\documentclass[12pt]{article}

\usepackage{graphics}
\usepackage{amssymb}

\newcommand{\rf}[1]{(\ref{#1})}
\newcommand{\bea}{\begin{eqnarray}}
\newcommand{\eea}{\end{eqnarray}}

\newcommand{\e}{\mbox{e}}
\renewcommand{\d}{\mbox{d}}

\renewcommand{\L}{\Lambda}
\renewcommand{\b}{\beta}
\renewcommand{\a}{\alpha}

\renewcommand{\th}{\theta}
\newcommand{\ep}{\varepsilon}

\newcommand{\del}{\delta}
\newcommand{\Del}{\Delta}

\renewcommand{\k}{\kappa}
\newcommand{\oh}{\frac{1}{2}}
\newcommand{\oq}{\frac{1}{4}}

\newcommand{\ra}{\rangle}
\newcommand{\la}{\langle}

\newcommand{\mi}{\!-\!}
\newcommand{\equ}{\!=\!}
\newcommand{\pl}{\!+\!}

\def\void{}
\def\labelmark{}

\newenvironment{formula}[1]{\def\labelname{#1}
\ifx\void\labelname\def\junk{\begin{displaymath}}
\else\def\junk{\begin{equation}\label{\labelname}}\fi\junk}%
{\ifx\void\labelname\def\junk{\end{displaymath}}
\else\def\junk{\end{equation}}\fi\junk\labelmark\def\labelname{}}

{\ifx\void\labelname\def\junk{\end{array}\end{displaymath}}
\else\def\junk{\end{array}\right.\end{equation}}
\fi\junk\labelmark\def\labelname{}\def\junk{}
}

\newcommand{\beq}{\begin{formula}}
\newcommand{\eeq}{\end{formula}}
\newcommand{\beqv}{\begin{formula}{}}

\begin{document}

\hfill AEI-2000-074

\vspace{.1cm}

\hfill 29 Nov 2000

\begin{center}
\vspace{24pt}
{ \large \bf Non-perturbative 3d Lorentzian Quantum Gravity}

\vspace{30pt}

{\sl J. Ambj\o rn}$\,^{a}$,
{\sl J. Jurkiewicz}$\,^{b}$ and
{\sl R. Loll}$\,^{c}$

\vspace{24pt}

$^a$~The Niels Bohr Institute, \\
Blegdamsvej 17, DK-2100 Copenhagen \O , Denmark\\
{\it email: ambjorn@nbi.dk}

\vspace{10pt}
$^b$~Institute of Physics,\\
Jagellonian University,\\
Reymonta 4, PL 30-059 Krakow, Poland\\
{\it email: jurkiewi@thrisc.if.uj.edu.pl}

\vspace{10pt}
$^c$~Albert-Einstein-Institut,\\
Max-Planck-Institut f\"{u}r Gravitationsphysik,\\
Am M\"uhlenberg 1, D-14476 Golm, Germany\\
{\it email: loll@aei-potsdam.mpg.de}

\vspace{48pt}

\end{center}

%\addtolength{\baselineskip}{0.20\baselineskip}
%\vspace{2cm}

\begin{center}
{\bf Abstract}
\end{center}

We have recently introduced a
discrete model of Lorentzian quantum gravity,
given as a regularized non-perturbative state sum 
over simplicial Lorentzian space-times, each possessing
a unique Wick rotation to Euclidean signature. 
We investigate here the phase structure of the Wick-rotated path
integral in three dimensions with the aid of computer simulations.
After fine-tuning the cosmological constant to its critical
value, we find a whole range of the gravitational coupling
constant $k_{0}$ for which the functional integral is dominated by
non-degenerate three-dimensional space-times. We 
therefore have a situation in which a well-defined ground state
of extended geometry is generated dynamically from a non-perturbative 
state sum of fluctuating geometries.
Remarkably, its macroscopic scaling properties
resemble those of a semi-classical spherical universe.
Measurements so far indicate that $k_{0}$ defines an overall
scale in this extended phase, without affecting the physics of
the continuum limit.
These findings provide further evidence that discrete {\it Lorentzian} 
gravity is a promising candidate for a non-trivial theory of
quantum gravity.

\vspace{12pt}
\noindent

%\vfill

\newpage

\section{Introduction}

The aim of discrete approaches to quantum gravity is
the non-perturbative construction of a quantum theory
of gravity as the continuum limit of a discretized state sum,
analogous to Feynman's construction of the quantum-mechanical
propagator of a particle \cite{iblecturenotes}. 
Among quantizers of gravity, path-integral formulations 
have fallen into disrepute,
both because of the non-renormalizability 
of the perturbation series and because of the unboundedness of the action,
which seems to render Euclidean approaches (at least formal
continuum path integrals and covariantly formulated 
cosmological models) ill-defined. 
Although this does not necessarily constitute an obstacle to the 
existence of a {\it non-perturbative}
path integral, it raises the question
of how such a quantity is to be constructed.

There is little hope of evaluating the continuum path integral
directly, because of the complicated functional form of the
gravitational action and because of quantum-field theoretic 
divergences, with the ensuing need to regularize in a way
compatible with the symmetries of the theory.
This leads to additional complications, since the 
invariance group of general relativity is the group of space-time 
diffeomorphisms, and not the Poincar\'e group of 
quantum field theory on a fixed, flat background, so that
standard regularization methods cannot be applied.

There are alternative regularization schemes, much in the
spirit of Feynman's treatment of the non-relativistic particle,
formulated on discretized versions of the space of all
space-time geometries (that is, of space-time metrics modulo
diffeomorphisms). 
One popular class of models is
based on applying Regge's idea \cite{regge}
of approximating smooth space-times 
$(M,g_{\mu\nu})$ by
piecewise linear, simplicial manifolds in a quantized context.
Unfortunately such models have to date produced little convincing
evidence of interesting continuum physics, for reasons
that are ultimately not well understood (see \cite{4dreview} for a 
recent review of discrete models in 4d).  

A serious criticism one can level at these models is that they
are all formulated for positive-definite Euclidean (Riemannian) metrics
$g_{\mu\nu}^{\rm Eu}$,
and not for physical metrics $g_{\mu\nu}^{\rm Lor}$ 
of indefinite, Lorentzian signature. 
This is done for technical reasons, since in a concrete regularized
formulation one must make sure that the state sum/integral $Z$
{\it converges}. However, it is important to realize that in a
{\it non}-perturbative context 
\beq{subst}
Z^{\rm Lor}=\int\limits_{\frac{{\rm Lor}(M)}{{\rm Diff}(M)}} 
[{\cal D}g^{\rm Lor}_{\mu\nu}]
\e^{i S[g^{\rm Lor}]}\; \longmapsto \;
Z^{\rm Eu}=\int\limits_{\frac{{\rm Eu}(M)}{{\rm Diff}(M)}} 
[{\cal D}g^{\rm Eu}_{\mu\nu}]
\e^{- S[g^{\rm Eu}]}
\eeq
is an {\it ad hoc} substitution: away from a handful of
metrics with special symmetries (for example, flat Minkowski space),
there is no straightforward ``Wick rotation'' $g_{\mu\nu}^{\rm Lor}
\mapsto g_{\mu\nu}^{\rm Eu}$ (or, equivalently, $t\mapsto -it$). 

It may well be that the absence of 
an interesting continuum limit in these statistical models of
dynamical geometries is related to the absence of
any Lorentzian structure in their partition functions.
This observation has motivated us to construct a well-defined discrete
quantum gravity model in terms of {\it Lorentzian} geometries. 
A suitable
starting point is the method of {\it dynamical triangulations},
a variant of the quantum Regge calculus program, which has the
advantage of being amenable to both numerical simulations and
analytic treatments. Following Regge's original concept of describing
``geometry without coordinates'', the ``sum over all paths'' 
is performed directly over physically inequivalent geometries.
Unlike in continuum path-integral approaches, there is no 
need to introduce coordinates and to subsequently gauge-fix them.
(In this sense, diffeomorphism-invariance is manifest.)
In our model, instead of using equilateral Euclidean
triangulations, we take the state sum over a certain class of
Lorentzian Regge manifolds, obtained by gluing together a number 
of simple simplicial Lorentzian building blocks.

The model has been constructed explicitly in 2, 3 and 4 space-time
dimensions \cite{al,ajl,ajl2}, and been solved exactly in d=2
\cite{al,alnr}. 
Each Lorentzian geometry (or ``history'') has a foliated structure,
with a (discrete) proper time $t$ labelling successive spatial
slices. (Note that this proper time is simply one of the parameters 
characterizing the discrete geometries, 
and not a ``gauge choice'', since the formalism is completely
coordinate-invariant from the outset.) In addition, each history
has a causal structure, induced from the piecewise linear 
Lorentzian metric structure. Each spatial slice is a $(d\mi 1)$-dimensional 
equilateral triangulation of Euclidean simplices, with squared
edge lengths $l_{\rm space}^2 \equ a^2$, and spatial topology
changes are not allowed (in line with the continuum notion of
causality).

A unique Wick rotation is defined 
on every Lorentzian history. It maps a given
triangulation with certain assignments of edge
lengths into the same topological triangulation, but with
the (squared) edge lengths of its {\it time-like} links 
(which interpolate between the spatial slices) redefined from
$l_{\rm time}^2 \equ -\alpha a^2$ to $l_{\rm time}^2 \equ +\alpha 
a^2$, where $\alpha >0$. This leads to an analytic continuation
from Lorentzian to Euclidean signature of the Regge action in the 
complex $\alpha$-plane and gives a {\it precise} meaning to the map 
(\ref{subst}).

For finite (discrete) volume, the Lorentzian gravity models thus 
obtained are
well-defined in the sense of being statistical systems whose
transfer matrix is bounded and positive. 
One is then interested in whether they exhibit any critical behaviour
as functions of the bare coupling constants, leading
to continuum theories of quantum gravity. Secondly, one wants
to compare their properties with those of the corresponding
Euclidean dynamically triangulated models. It should perhaps be
emphasized at this point that our non-perturbative 
path-integral method is in principle exact (and not {\it formal}).
It does involve a specific choice of a ``measure'', but one
would expect from universality arguments that the fine details of
any choices made at the discretized level will not alter the
continuum theory. 
 
As already mentioned above, in d=2 the Lorentzian model is exactly
soluble and lies in a different universality class from its
Euclidean counterpart (which can also be solved exactly and 
is better known as 2d Liouville quantum gravity). Its geometric
properties are different, which also affects its behaviour when
matter is coupled to the gravitational degrees of freedom 
\cite{aal1,aal2}. 
Quite remarkably, and unlike Liouville gravity, the coupled
system remains well-behaved beyond the so-called ``c=1 barrier''
(in our case, this is equivalent to the number of coupled Ising models 
exceeding 2). 

These are very interesting results from the point of view of
systems of two-dimensional random geometries, but our ultimate
interest lies in the physical, four-dimensional case, and the
physics of ``general relativity'' in 2, 3 and 4 dimensions is
very different. 
Where dynamically triangulated {\it Euclidean} models seem to go wrong
in $d>2$ is in the dominance of highly degenerate geometries over
their statistical ensembles. It is encouraging that one can
show the absence of the same type of geometries from the Lorentzian 
ensemble \cite{ajl,ajl2}, but one could still be worried about the
occurrence of (less extreme) pathologies. The only way to determine
whether Lorentzian gravity does indeed solve the problems of 
the Euclidean approach, is to investigate its
phase structure in the continuum limit,
either by numerical simulations or by solving it explicitly.

Before embarking on the physically relevant case of $d=4$, we will
in this paper investigate Lorentzian quantum gravity in three
dimensions. (Some of the results presented here have been announced
recently in \cite{amb2,plenary}.)
Apart from being a new statistical model of
three-dimensional fluctuating geometries, this also has
some interest from the point of view of quantum gravity proper. 
Although largely an unphysical
theory, 3d quantum gravity is an extensively studied system 
\cite{3dgrav,carlip}.
It is often invoked as a model system for the full theory,
since its classical equations resemble in many ways those
of general relativity. The big difference from d=4 is the fact
that there are no propagating physical field degrees of freedom. 
After solving the constraints of the theory, only a finite-dimensional 
phase space remains. Although one has not yet been able to make full
use of this observation in a configuration space path-integral
formulation, it suggests that one may still be able to solve
3d gravitational models {\it analytically}.

The derivation of the partition function and the explicit construction 
of 3d Lorentzian simplicial space-times
was given in \cite{ajl}. In order to make this article self-contained,
we will summarize the main results below and at the
beginning of Sec.\ 3. 
The Einstein action of a given (smooth) Lorentzian
geometry in three dimensions is  
\beq{action}
S [g_{\mu\nu}] = \frac{1}{16\pi G} 
\int d^3x\; \sqrt{-\det g}\, (R-2\Lambda),
\eeq  
where $G$ and $\Lambda$ denote the gravitational  
and cosmological constants.
For the continuous, piece-wise linear geometries
employed in our simplicial discretization, 
we use the Regge form of the action \cite{regge,sorkin}, 
expressed purely in terms of geometric (coordinate-independent) 
data, namely, the geodesic edge lengths of the simplicial complexes. 

In order to make the state sum well defined, 
we analytically continue the Regge action
associated with each 3d Lorentzian triangulation to
Euclidean form by changing the length assignments of all
time-like edges from $l_{\rm time}^2\equ\ - a^2$ to
$l_{\rm time}^2\equ + a^2$. 
The main aim of this article is to analyze the phase structure 
of the model defined by the path integral obtained after this
``Wick-rotation'', 
\beq{wick2}
\sum_{{\cal T}_{T}(S^1\times S^2)} 
\frac{1}{C(T)}\ \e^{i 
S(N_0,N_3,T)}\;\stackrel{l_{\rm time}^2\to
-l_{\rm time}^2}{\longmapsto}
\sum_{{\cal T}_{T}(S^1\times S^2)} 
\frac{1}{C(T)}\ \e^{- 
S_{\rm E}(N_0,N_3,T)},
\eeq
where $C(T)$ is the order of the symmetry group of the 
triangulation $T$. The ``measure factor'' $1/C(T)$ appears 
naturally in the counting of unlabelled triangulations 
\cite{david}. As usual in the theory of critical phenomena,
we do not expect the detailed choice of the measure to affect
the continuum limit of the theory, a behaviour that has already been
corroborated by the 2d models of Euclidean and Lorentzian
quantum gravity.
In (\ref{wick2}), the Euclidean three-dimensional Regge action 
$S_{\rm E}$ is 
expressed in terms of the 
total numbers of vertices and tetrahedra, 
$N_0$ and $N_3$, according to
\beq{euact2}
S_{\rm E}(N_0,N_3,T) = -k_0 N_0 +k_3 N_3,
\eeq
with the associated dimensionless bare coupling constants
\beq{couplings}
k_0 = \frac{a}{4G}, ~~~~
k_3= \frac{a^3\Lambda }{48\sqrt{2}\pi G} +\frac{a}{4G} (3\k-1)
\eeq
(see appendix 1 for a derivation).
This form of the action is familiar from former work in 
Euclidean dynamical triangulations \cite{former}. 
The parameter $k_0$ is proportional to 
the bare inverse gravitational coupling constant, while $k_3$ is a 
combination of the bare gravitational and cosmological constants
(often referred to -- somewhat imprecisely -- simply as the
(bare) cosmological constant). 
The geodesic ``lattice spacing'' (edge length) is given by $a >0$ 
and $\k\pi = \arccos(1/3)$ is the dihedral angle of an
equilateral tetrahedron.

To keep things simple, we are assuming that the spatial 
slices have the topology of two-spheres.
In addition, for the convenience of the numerical simulations, 
we are using
periodic boundary conditions in the (Euclidean) time direction,
unless specified otherwise. 
The sum in \rf{wick2} is taken over the set of all causal
triangulations ${\cal T}_{T}(S^1\times S^2)$
compatible with this topology, and 
constructed according to the rules described in Sec.\ 2 below. 
The integer parameter $T$
denotes the total extent in (discrete) proper time,
i.e. the number of spatial slices of constant 
$t\in [0,T]$.\footnote{We have
slightly changed our notation with respect to \cite{ajl}, where the
{\it total} proper time was called $t$.}

In this article we explore the phase diagram of three-dimensional
discrete Lo\-rentz\-ian gravity.
We are particularly interested 
in identifying those regions of coupling-constant space 
where a continuum limit may exist. 
This is done with the help of Monte Carlo simulations 
of the statistical ensemble defined by \rf{wick2}, combined
with qualitative analytical arguments. 

The rest of this article is 
organized as follows: the next section contains some general
considerations on the behaviour of discrete quantum gravity
models under renormalization. 
In Sec.\ 3 we describe the implementation of the Monte Carlo
algorithm 
on the ensemble ${\cal T}_{T}(S^1\times S^2)$ of
causal 3d geometries, 
as well as a characterization of the triangulations
and the Monte Carlo moves in terms of dual graphs.
Our numerical results are presented in Sec.\ 4. We characterize 
the different phases by measuring various
geometric observables, and 
give a detailed description of the geometric
properties of the physically interesting ``extended'' phase.
Finally, Sec.\ 5 contains a summary and discussion of our results.
There are three appendices. In App.\ 1, various identities
and derivations for 3d simplicial geometries are collected,
App.\ 2 contains more details on dual graphs, and App.\ 3
some technical specifications of the Monte Carlo simulation.

\section{Renormalization in discrete quantum gravity}

As we know from the study of Euclidean simplicial
quantum gravity \cite{former}, there is a well-defined 
strategy to search for possible 
continuum limits for the type of discretized gravity model 
we are considering.  For each 
value $k_0$ of the bare inverse gravitational 
coupling there is a critical value $k_3^c(k_0)$ of the
bare cosmological constant
such that the model is well defined for $k_3 >k_3^c(k_0)$ and 
diverges for $k_3 < k_3^c(k_0)$. One can hope to obtain
a continuum limit for $k_3 \to k_3^c(k_0)$ because in this limit 
the expectation value
$\la N_3^n\ra$ may diverge for suitable powers $n$. 

This program has been carried out successfully in both 
Euclidean and Lorentzian quantum gravity
in d=2 (see \cite{rew2d} for a recent review).
It offers a non-perturbative field-theoretical definition 
of 2d quantum gravity where the bare 
cosmological constant $k_2$ is additively renormalized according to
\beq{2.1}
k_2 = k_2^c + \L a^2,
\eeq
with $\L$ denoting the 2d continuum cosmological constant, and where 
the critical $k_2^c$ comes entirely from the entropy of the 
two-dimensional triangulations. 
We expect an analogous additive renormalization of the bare cosmological
coupling constant $k_3$ in 3d quantum gravity,
but in this case $k_3^c$
depends not only on the entropy of the three-dimensional 
triangulations but also on the gravitational 
coupling $k_0$, since the Einstein action gives 
a non-trivial weight to each triangulation (contrary to 
two dimensions, where the curvature term is topological).

Taking the infinite-volume limit of a regularized quantum field 
theory does not necessarily 
lead to a continuum quantum field theory. 
For example, the Ising model on a 
infinite two-dimensional lattice will only represent a $c\equ 1/2$
conformal field theory if at the same time
the temperature (which plays the role
of a coupling constant in the theory) is fine-tuned 
to the critical temperature of the Ising model. 
Only when approaching the 
critical temperature will the long-range spin fluctuations become 
important and allow us to forget about the details of the lattice
regularization, thereby
making contact with continuum physics. 

By contrast, 2d quantum gravity 
is an example where the infinite-volume 
limit of the regularized theory automatically leads to the 
continuum theory. 
This was to be expected as the cosmological coupling is 
the only coupling constant of the theory, and at the
same time conjugate to the space-time volume.

It is not immediately clear what to expect in three-dimensional 
quantum gravity. The classical theory (after gauge-fixing) has no
propagating {\it field} degrees of freedom, but is described by
a finite number of (Teichm\"uller) parameters, whose number depends on the
topology of the spatial slices. 
Since in the theory of critical 
phenomena a divergent correlation length and the 
associated fine-tuning of a coupling constant are usually associated
with a {\it field} degree of freedom, it is tempting to conjecture
that the situation will be as in two dimensions, namely,
that the infinite-volume limit of 3d Lorentzian quantum gravity 
(obtained by fine-tuning the cosmological constant) coincides
with the continuum limit, without 
the need for further fine-tuning. 

In three dimensions, we must in addition understand which
role the gravitational coupling constant plays in our formulation.
In the exponentiated action, it multiplies the curvature
term $\int \d^3 x \,\sqrt{\det g(x)} \, R(x)$ of the classical
Einstein action, from which the classical dynamics is derived. 
Moreover, it is exactly this term that gives rise to the
non-renormalizability of three-dimensional quantum gravity,
when one considers perturbation theory around a classical solution. 
This means that it does not make much sense to expand around 
a given flat background in a conventional way. Although
the underlying quantum theory may not have any divergences
(since there may not be any propagating field degrees of 
freedom), we are likely to end up with a infinite 
set of divergent counterterms\footnote{One could try to view 
such a theory as an effective low-energy theory with limited 
predictive power, 
much in the same way as the non-linear sigma model is used as an effective 
field theory for pion physics, describing some aspects 
of low-energy QCD. However, this is not what we 
are after in a non-perturbative definition of 3d quantum 
gravity. We want a theory whose predictions in the 
continuum limit are {\it all} independent of the cut-off of the 
regularized theory, although they may in principle depend 
on a non-perturbatively induced mass scale.}, unless there is
some as yet undiscovered cancellation mechanism. 

There are well-known ways 
to circumvent this deadlock in the case of three-dimensional gravity, 
for example, by quantizing in the 
reduced, finite-dimensional phase space, either in a geometric formulation
using $g_{\mu\nu}$ or using gauge-theoretic (Chern-Simons) variables
\cite{3dgrav,carlip}. (However, it should be remembered that even 
classically, the 
relation between these two ``time-full'' and ``time-less'' 
formulations is only partially understood \cite{mon1}.)
How this is reflected in a path-integral quantization in terms
of geometries is much less clear (we mean here a {\it configuration}
space path integral, rather than a phase space path integral). 
As far as we know, there is not even a generally accepted
answer to whether or not the gravitational coupling constant should be
renormalized.

If our discretized non-perturbative model possesses a continuum limit, 
it should provide an answer to this question. 
For example, the presence
of a non-trivial second-order phase transition at a specific 
value of $k_0$ would strongly suggest to
take the continuum limit by fine-tuning $k_0$ to this 
point, defining in the process the renormalization of the 
gravitational constant. The issue of non-renormalizability 
could be circumvented if the fixed point 
was non-trivial, not allowing for a simple expansion in geometry. 
To some extent this is realized in $2+\ep$-dimensional
quantum gravity which possesses a non-trivial fixed point
\cite{kawai1,kawai2,kawai3}. Of course the challenge in such a scenario 
would be to understand how its excitations are related 
to the degrees of freedom, or rather the lack of degrees of freedom,
of the classical 3d gravity theory. Conversely, if no 
second- or higher-order transition is present and one can still 
define a continuum limit, it is likely that no renormalization 
of the gravitational constant is necessary.

In order to find answers to these questions, we will in
this article analyze data coming from numerical studies
of three-dimensional Lorentzian gravity. Attempts to solve the
model analytically are under way and will be reported elsewhere
\cite{toappear}.
As will be described in the following, our investigation
provides evidence that 
\begin{itemize}
\item[(i)] a continuum limit exists;
\item[(ii)] there is a well-defined ground state which dominates
the functional integral and thus represents a ``background geometry'';
\item[(iii)] the bare gravitational coupling constant sets
a length scale for the geometry, but is not renormalized.
\end{itemize}

\section{Numerical implementation of the model} 

\begin{figure}[t]
%\vspace{-5cm}
\centerline{\scalebox{0.6}{\rotatebox{0}{\includegraphics{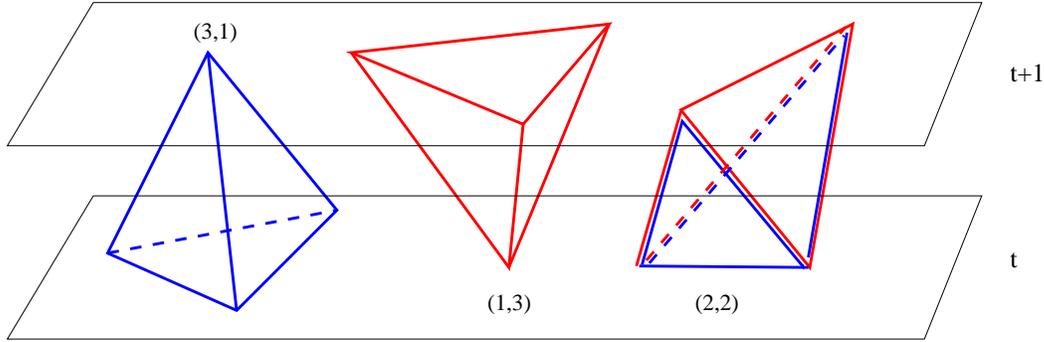}}}}
%\vspace{-5cm}
\caption[fig1]{The three types of tetrahedral building blocks used in
discrete 3d Lorentzian quantum gravity, and their location with respect
to the spatial slices of constant integer $t$.}
\label{fig1}
\end{figure}

Which are the three-dimensional Lorentzian geometries contributing
to the state sum \rf{subst}?
Starting from a sequence of two-dimensional equilateral triangulations,
a three-dimensional Lorentzian triangulation is obtained by filling
the spaces between pairs of such spatial slices by three types of
tetrahedral building blocks, in such a way as to form a 
{\it simplicial manifold}.
They are (i) the so-called (3,1)-tetrahedra with a 
triangle in the spatial $t$-plane and a vertex in the spatial
$t\pl 1$-plane; their number in any given sandwich $[t,t+1]$ is denoted
by $N_{31}(t)$, and their total number by $N_{31}$; 
(ii) the (1,3)-tetrahedra 
with a vertex in the $t$-plane and a triangle in the $t\pl 1$-plane; 
their number in any given sandwich $[t,t+1]$ is denoted
by $N_{13}(t)$, and their total number by $N_{13}$; 
and (iii) the (2,2)-tetrahedra with one link in the $t$-plane 
and another one in the $t\pl 1$-plane;
their number in any given sandwich $[t,t+1]$ is denoted
by $N_{22}(t)$, and their total number by $N_{22}$
(see Fig.\ \ref{fig1}).

Several of our numerical measurements involve the two-volume of 
spatial slices. In our model there are two natural ways of defining the
spatial volume at time $t$. One can define it simply as the number
of triangles in the spatial slice of constant integer $t$, 
\beq{spat1}
N_{2}^{(s)}(t)\equiv N_{31}(t) \equiv N_{13}(t-1)
\eeq
or measure it at half-integer $t$ and define\footnote{This
definition is the simplest one in that it counts the number of
building blocks at $t=1/2$. We could put in individual
weights reflecting the actual areas of the cross-sections
of the tetrahedra, but it would not affect our results below.}
\beq{spat2}
N_{2}^{(s+1/2)}(t):= N_{13}(t)+N_{31}(t)+N_{22}(t).
\eeq 
In a phase of extended geometry (such that $N_{13}\sim N_{31}\sim
N_{22}$), both definitions should lead to equivalent results.
For most purposes, we have found it convenient to work with
the two-volumes $N_{2}^{(s+1/2)}(t)$.

We will explore the infinite-volume limit of the ensemble of discrete
Lorentzian geometries by performing a Monte Carlo simulation where each
suggested local change of triangulation (a ``move'') is accepted 
or rejected according to certain probabilities depending on the 
change in the action and the local geometry.

Our local updating algorithm consists of five basic moves.
They change
one Lorentzian triangulation 
into another, while preserving the constant-time slice structure, as 
well as the total proper time $T$. 
We are confident that this set of moves is ergodic in the space of all
allowed Lorentzian triangulations at fixed $T$, although we do not as yet
have a complete formal proof. Note that all of the moves described 
below will
be rejected in the updates if they lead to triangulations where 
pairs of
vertices are connected by more than one link or where triplets of vertices
belong to more than one triangle, since this violates the
{\it simplicial manifold} property.
Let us now describe each of the moves in turn:
\begin{itemize}
\item[(1):] Consider two neighbouring triangles in the spatial $t$-plane.
Each of them belongs to a tetrahedron above and below that plane.
Assume now that both the two (3,1)-tetrahedra above and the two
(1,3)-tetrahedra below share a triangle.
Together, the four tetrahedra form a diamond whose tips lie
in the $t\mi 1$- and in the $t\pl 1$-plane, and whose intersection
with the $t$-plane is a square.
The move consists in flipping the link that forms the diagonal of
this square to the opposite diagonal, accompanied by the corresponding
reassignment of the tetrahedra constituting
the diamond (Fig.\ \ref{fig2}). 
\begin{figure}[t]
%\vspace{-5cm}
\centerline{\scalebox{0.6}{\rotatebox{0}{\includegraphics{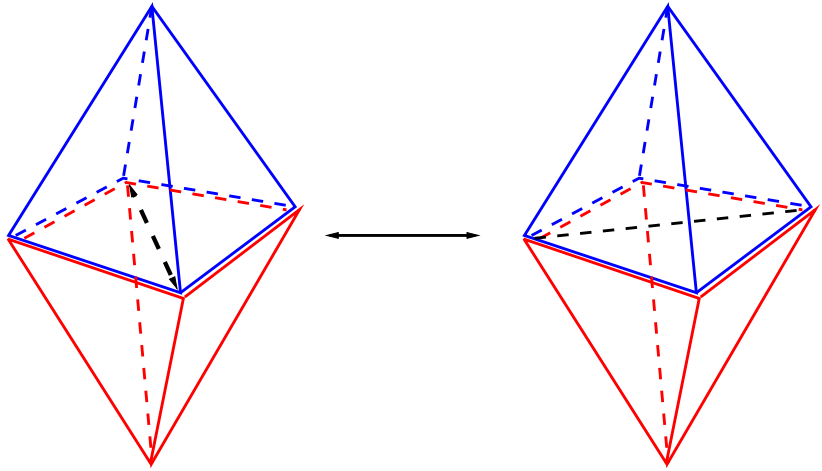}}}}
%\vspace{-5cm}
\caption[fig2]{In a flip move, the four tetrahedra inside a diamond
configuration are reassigned by flipping the diagonal of the
central square.}
\label{fig2}
\end{figure}
\item[(2\&3):] Consider a triangle in the $t$-plane together with its 
two neighbouring tetrahedra, whose two tips 
$v_{t\pl 1}$ and $v_{t\mi 1}$ lie in the $t\pl 1$- and 
the $t \mi 1$-plane. We can always insert a vertex $v_t$
at the centre of the triangle and connect it to the exterior
vertices of this configuration by adding five internal links, thus
replacing the original two tetrahedra by six (Fig.\ \ref{fig3}).
The corresponding {\it inverse move} can only be performed if we can 
identify a vertex $v_t$ of order six (i.e. belonging to six
tetrahedra), together with two links $(v_t,v_{t\mi 1})$ and
$(v_t,v_{t\pl 1})$ which are both of order three. In this 
case one can just remove $v_t$ and both links $(v_t,v_{t\pm 1})$, 
replacing in an obvious way the six tetrahedra by two.
\begin{figure}[t]
%\vspace{-5cm}
\centerline{\scalebox{0.6}{\rotatebox{0}{\includegraphics{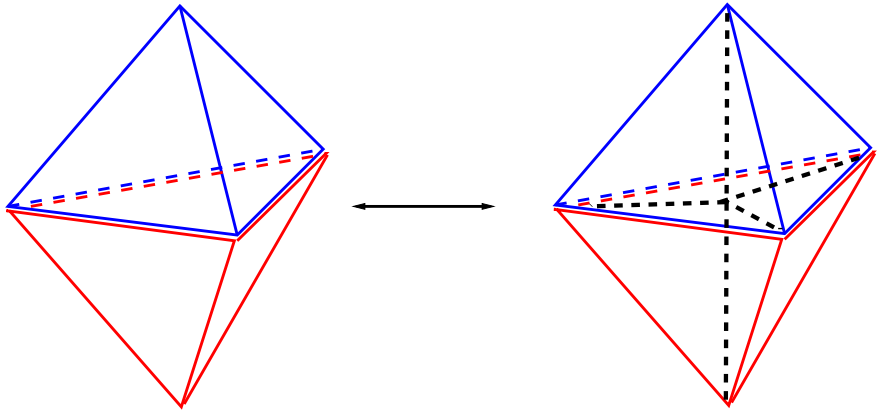}}}}
%\vspace{-5cm}
\caption[fig3]{Insertion or removal of a vertex in the central
triangle transforms two into six tetrahedra and vice versa.}
\label{fig3}
\end{figure}
\item[(4\&5):] The fourth move can be performed on any configuration
consisting of a pair of a (2,2)- and a (3,1)- (or a (1,3)-) tetrahedron 
having a triangle in common.  
We can remove the triangle (but not its links and vertices)
and insert a link dual to it, 
connecting the two vertices which did not belong to 
the triangle (see Fig.\ \ref{fig4}). 
In this way the original (3,1)- and (2,2)-tetrahedra are replaced
by one (3,1)- and two (2,2)-tetrahedra, without introducing any
changes in the two-dimensional spatial slices. 
The fifth move is the inverse of the fourth move, replacing a suitable
configuration of one 
(3,1)- (or (1,3)-)tetrahedron and two adjacent (2,2)-tetrahedra
by a pair of a (3,1)- (or (1,3)-) and a (2,2)-tetrahedron.
\begin{figure}[t]
%\vspace{-5cm}
\centerline{\scalebox{0.6}{\rotatebox{0}{\includegraphics{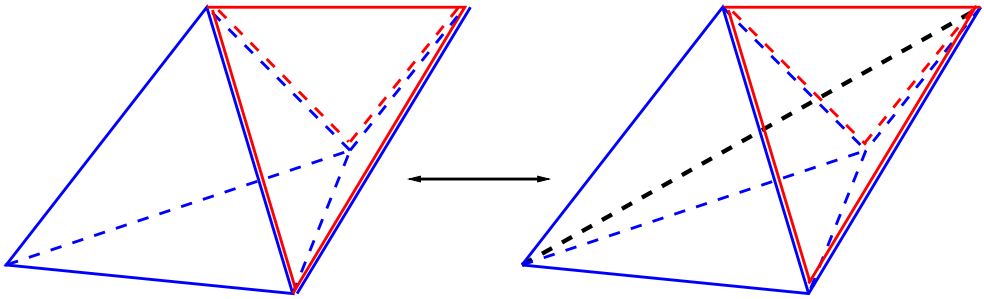}}}}
%\vspace{-5cm}
\caption[fig4]{A neighbouring pair of a (3,1)- and a (2,2)-tetrahedron
is replaced by another (3,1)-tetrahedron (whose tip lies now in the
top right-hand corner) and a (2,2)-tetrahedron on either of its
flanks.}
\label{fig4}
\end{figure}
\end{itemize}

Note that not all of the local moves preserve the three-volume. 
We will use 
a standard way of dealing with this situation, developped for 
dynamically triangulated models in dimensions three and four \cite{former}.
This method ensures that the volume of the system 
fluctuates around a prescribed 
value $N_3$, with a well-defined range of fluctuations.

\begin{figure}[t]
%\vspace{-5cm}
\centerline{\scalebox{0.7}{\rotatebox{0}{\includegraphics{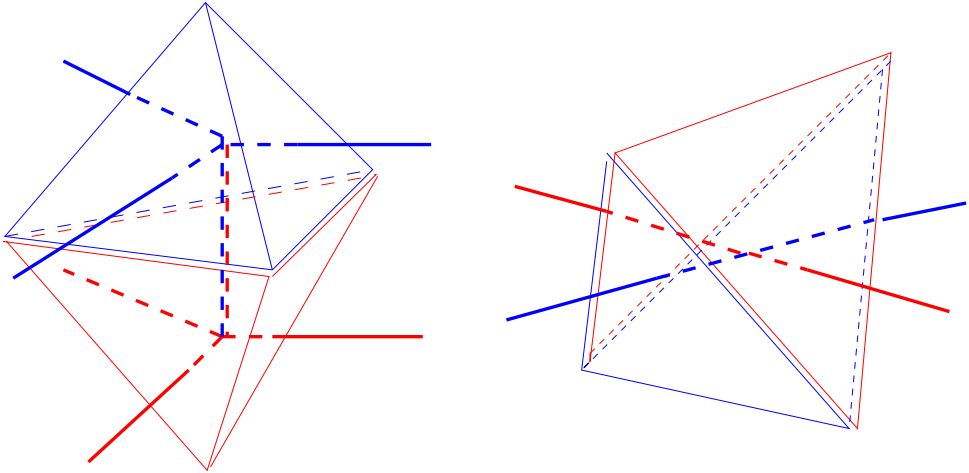}}}}
%\vspace{-5cm}
\caption[fig5]{How the tetrahedral building blocks give rise
to dual bi-coloured graphs (drawn as red and blue horizontal
lines) at half-integer $t$.}
\label{fig5}
\end{figure}

In the implementation of the numerical code it is
convenient to work not with the triangulations and their constituents
but with the dual graphs, which are given by specific classes of 
$\phi^4$-graphs. Like the triangulations, all of the graphs 
have a foliated structure.
This is most naturally associated with half-integer
times, because the vertices of the dual graph are located at the 
centres of the tetrahedra of the original triangulation.
To visualize the geometry of the gluings and the moves in this
dual language, we adopt a colouring for the dual graphs.
A link dual to a triangle of a (3,1)-tetrahedron is ``blue'', 
and one dual to a triangle of a (1,3)-tetrahedron is ``red''.
(We have already anticipated this in Figs.\
\ref{fig1}-\ref{fig4} by giving each {\it triangle} a definite
colour; only the links affected by the Monte-Carlo moves are
drawn in black.)
This results in a unique colouring for all links dual to 
``time-like'' triangles (lying {\it in between} spatial slices),
since it is not possible to glue directly a red to a blue triangle.
This can only be done if the triangles are space-like (i.e. if they
are both contained in the same slice $t\equ const$). The links dual to
such triangles are therefore double-coloured (Fig.\ \ref{fig5}, left).
We can now construct for each ``sandwich'', i.e. each triangulated
space-time slice $[t,t\pl 1]$, a bi-coloured graph (with 
topology $S^2$) by
projecting all {\it uni-coloured} dual links associated with
the sandwich to the plane at $t\pl 1/2$. 

In this way each (3,1)-tetrahedron gives rise to three blue links,
sharing a trivalent intersection. Each of the links can end at the 
centre of either another (3,1)-tetrahedron or 
a (2,2)-tetrahedron, but never at the centre of a (1,3)-tetrahedron.
(An analogous statement holds for the triplet of red
links associated with a (1,3)-tetrahedron.)
Consequently, each (2,2)-tetrahedron in the sandwich corresponds to
a four-valent vertex of the dual graph, with alternate colours
blue-red-blue-red for the incoming links
(Fig.\ \ref{fig5}, right).
The end result is a combined red-and-blue graph in the 
$t\pl 1/2$-plane. Moreover, each such graph occurs 
in the large-$n$ limit of the perturbative expansion of the 
two-matrix model defined by the partition function
\beq{matrix}
Z(\alpha_R,\alpha_B,\beta)=
\int \d \phi_R \,\d\phi_B \;\e^{n \,{\rm tr}\, [-\oh (\phi_R^2+\phi_B^2)
+\frac{\a_R}{3} \phi^3_R + \frac{\a_B}{3} \phi^3_B+ \frac{\b}{2} 
\phi_R\phi_B \phi_R\phi_B]},
\eeq
where, as usual, the quadratic terms give rise to propagators or
links, and the cubic and quartic interaction terms correspond exactly
to the tri- and four-valent intersections illustrated in
Fig.\ \ref{fig5}. Note that not all graphs generated by (\ref{matrix})
correspond to allowed Lorentzian triangulations, since their duals
may violate the 3d simplicial manifold constraints.
This matrix model (with some additional assumptions about universality) 
can be taken as the starting point for an analytical solution 
of the transfer matrix of simplicial 3d quantum gravity \cite{toappear}.

The time-evolution in the dual picture can be thought of as follows.
A bi-coloured graph at time $t\pl 1/2$ consists of two components:
a blue $\phi^3$-graph dual to the triangulation
at time $t$, and a red $\phi^3$-graph dual to the triangulation at 
$t\pl 1$. The way in which the two original spatial triangulations
are glued together is encoded in the intersection pattern of
the ``superposition'' of the two graphs at time $t\pl 1/2$.
If we view the blue and red trivalent graphs as representing
in- and out-states, their transition amplitude is a function of
the number of topologically inequivalent ways of superposing the 
two graphs (subject to some ``dual'' manifold constraints --
see App.\ 2 for details).

\begin{figure}[t]
%\vspace{-5cm}
\centerline{\scalebox{0.6}{\rotatebox{0}{\includegraphics{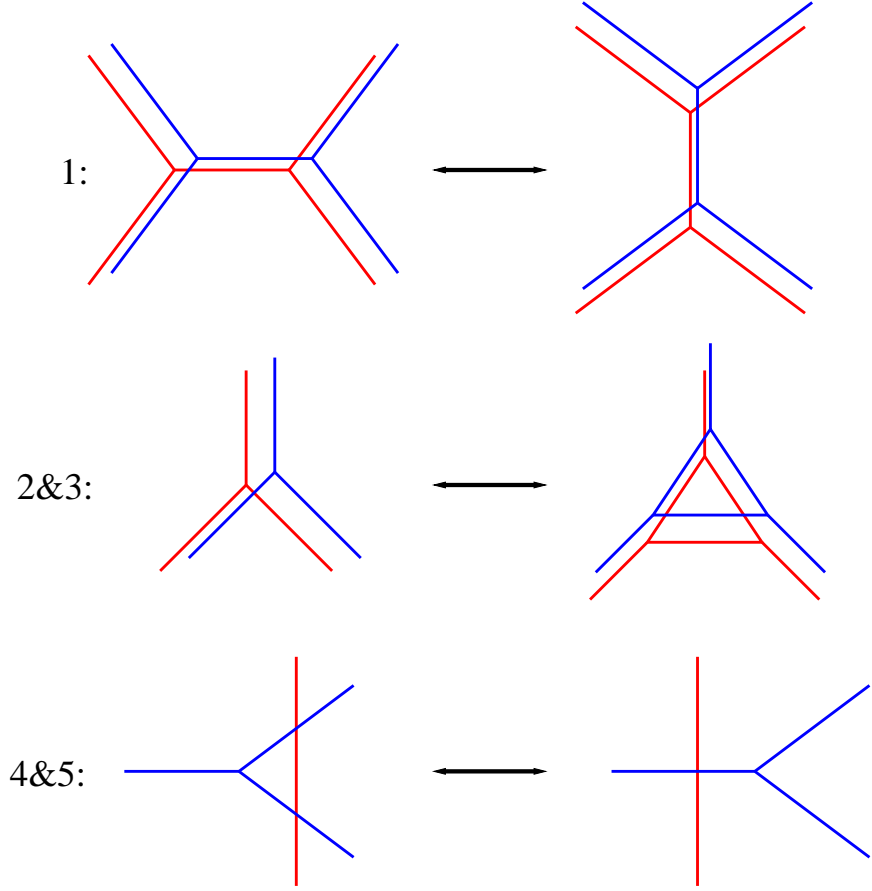}}}}
%\vspace{-5cm}
\caption[fig5a]{Graphical illustration of the Monte Carlo moves
in terms of dual bi-coloured graphs. In moves 1 and 2\&3, dual links
from two adjacent sandwiches are affected. Moves 4\&5 take place
within a given $[t,t\pl 1]$-sandwich.}
\label{fig5a}
\end{figure}

The five Monte Carlo moves described earlier can also be
rephrased in the language of intersecting
coloured $\phi^3$-graphs, as illustrated in Fig.\ \ref{fig5a}. 
The diagrams appearing in Fig.\ \ref{fig5a} contain all dual
links affected by a given move. Since the moves 1, 2 and 3 are
symmetric with respect to the plane $t\equ const$,
a change in one of the trivalent graphs is always accompanied
by an equivalent change of its mirror image of the opposite
colour. Moves 4 and 5 assume a particularly simple form: one
link of a given colour is ``dragged across'' a vertex of
the opposite colour. (Note that in the graphical representation of
this particular move, the cubic vertices represent 
(3,1)- or (1,3)-tetrahedra and the blue-red
crossings (2,2)-tetrahedra.)
Details about the numerical implementation in terms of this 
dual picture (including lattice sizes, update efficiency, 
number of sweeps etc.)
can be found in Apps.\ 2 and 3.

\section{Numerical results}

Having presented our numerical set-up, we are now in a position
to extract a number of physical properties of the Lorentzian
model. We will first investigate the phase diagram of the regularized 
theory, and try to understand which of the continuum-limit scenarios 
outlined in Sec.\ 2 is realized. We will then analyze the geometry
characterizing the different phases. 
Since we have a distinguished (and coordinate-invariant) notion of
proper time $t$, we can extract invariant information of the system by
studying correlation functions in $t$.
This will be done by measuring
distributions of spatial slice-volumes $N_{2}^{(s+1/2)}$ as a function
of the total proper time $T$ and correlators 
$\la N_{2}^{(s+1/2)}(t_1)N_{2}^{(s+1/2)}(t_2)\ra $ between spatial
volumes, as well as the intrinsic Hausdorff dimension $d_{H}^{\rm sp}$ 
of a typical spatial slice.

\subsection{The phase diagram}

\begin{figure}[t]
\vspace{-5cm}
\centerline{\scalebox{0.75}{\rotatebox{0}{\includegraphics{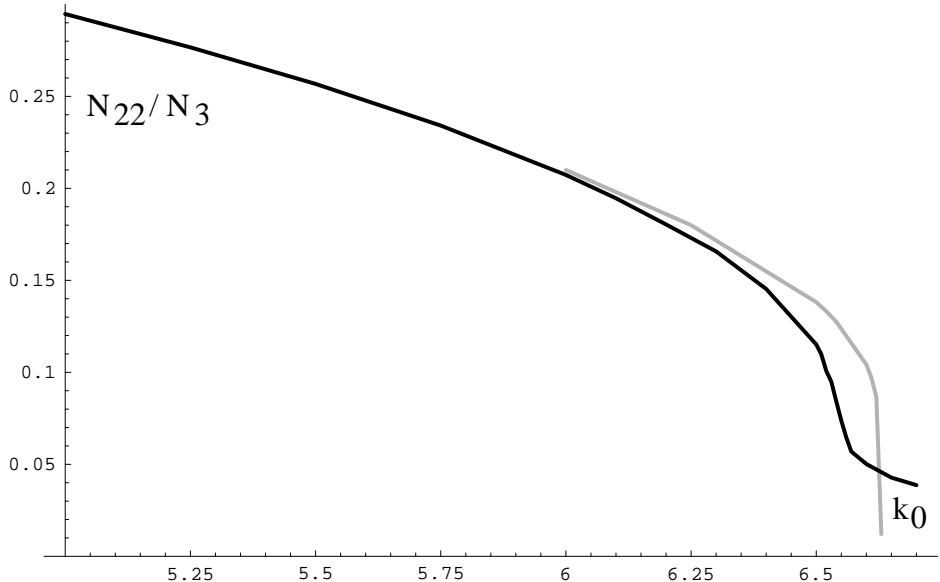}}}}
\vspace{-5cm}
\caption[fig6]{The order parameter $\tau=N_{22}/N_3$ for configurations
with $T\equ 64$, and $N_3\equ 16,000$ (long dark curve) and 
64,000 (short light curve), plotted as a function of $k_0$. 
The curve is a linear interpolation between data points.
(Error bars smaller than width of curve.)}
\label{fig6}
\end{figure}

\begin{figure}[t]
\vspace{-5cm}
\centerline{\scalebox{0.75}{\rotatebox{0}{\includegraphics{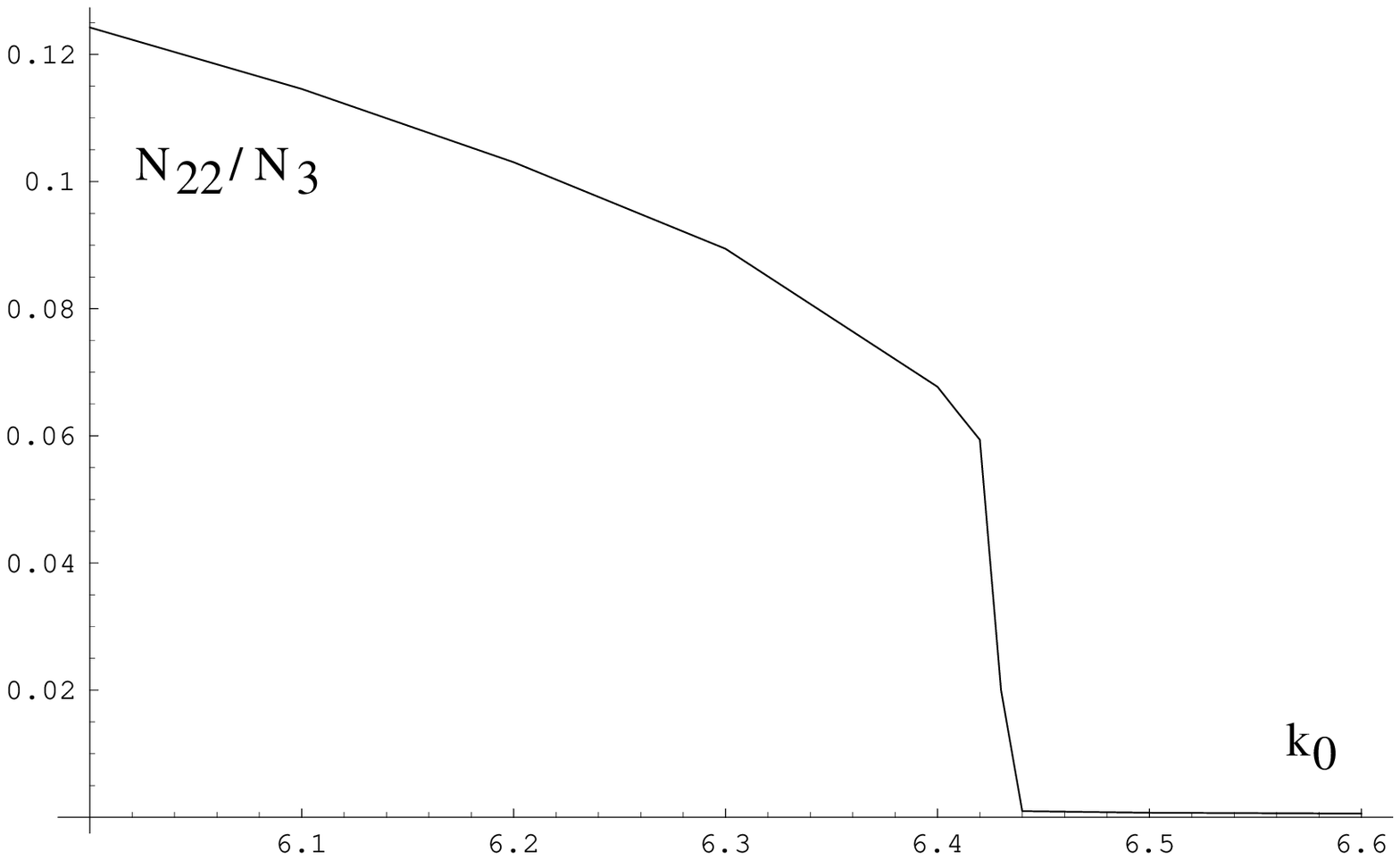}}}}
\vspace{-5cm}
\caption[fig7]{The order parameter $\tau=N_{22}/N_3$ for configurations
with $T \equ 1$ and $N_3\equ 16,000$, but with {\it free} 
boundary conditions, plotted as a function of $k_0$. 
The curve is a linear interpolation between data points.
(Error bars smaller than width of curve.)}
\label{fig7}
\end{figure}

In order to explore the phase diagram of the regularized Lorentzian
model we must find an order parameter, and explore how it 
changes with the coupling constant, in this case $k_0$.  
We have found that the ratio between the total number $N_{22}$ of 
(2,2)-tetrahedra and the total space-time volume $N_{3}$, 
\beq{tau}
\tau= \frac{N_{22}}{N_3}\equiv \frac{N_{22}}{N_{22}+N_{31}+N_{13}},
\eeq
serves as an efficient order parameter. 
We shall not be concerned with a continuum interpretation
of this parameter (which is not obvious) since we will 
go on to show that no continuum physics is associated with
the transition we observe as a function of $\tau$.
In Figs.\ \ref{fig6} and \ref{fig7}
we show the ratio $\tau$ as a function of $k_0$ for two different 
types of space-time configurations. In Fig.\ \ref{fig6} all geometries
have 64 spatial slices ($T\equ 64$), with total space-time volumes 
$N_3\equ 16,000$ and $N_3\equ 64,000$. One 
observes a rapid drop to zero of $\tau(k_0)$ around $k_0 \approx 6.64$.
Increasing $N_3$, the drop becomes a jump, characteristic for a 
first-order phase transition. A detailed study of the neighbourhood 
of $k_0 \equ 6.64$ reveals a (weak) 
hysteresis as one performs a cycle, moving above 
and below the critical value $k_0^c$, again as expected in a 
first-order transition.  
The location of the phase transition depends weakly on 
the total length $T$ in time-direction, and the jump in $\tau$
becomes more pronounced for larger volumes.  
Fig.\ \ref{fig7} is an illustration of the extreme 
case $T\equ 1$, where we have chosen {\it free} instead of periodic boundary
conditions (the triangulations at the initial 
time $t\equ 1$ and the final time $t\equ 2$
are allowed to fluctuate freely), so that the topology of space-time 
is changed to $S^2\times [0,1]$.\footnote{In this situation the 
critical point is changed from $k_0^c 
\equ 6.64$ to $k_0^c \equ 6.42$.} 

As can be read off from Fig.\ \ref{fig6}, for $k_0 > k_0^c \approx 6.64$ 
only a minimal number of (2,2)-tetrahedra is present. 
This can be understood by rewriting the action \rf{euact2}
to make the dependence on $N_{22}$ explicit. 
In appendix 1 we derive 
\beq{n2}
S_E = \frac{k_0}{4} N_{22} + \Big(k_3-\frac{k_0}{4}\Big)\, N_3- 
2k_0 \, T,
\eeq
which shows that for 
fixed $N_3$ and $T$ (and positive gravitational coupling $k_0$) 
a minimal $N_{22}$ corresponds to a minimum of the Euclidean action. 

The entropy of configurations
with $N_{22}$ different from its minimal value will in general 
ensure that the ratio $\tau$ is different from zero, even when $N_3
\to \infty$. However, since the number of such configurations for fixed 
$N_3$ grows at most exponentially with $N_3$, this leaves the
possibility that for sufficiently large $k_0$ the term 
$\e^{-k_0 N_{22}/4}$ will dominate over the entropy contribution
and trigger a phase transition to a phase 
with only a minimal number of (2,2)-tetrahedra, such that $\tau\equ 0$
in the continuum limit.

\begin{figure}[t]
\vspace{-4cm}
\centerline{\scalebox{0.7}{\rotatebox{0}{\includegraphics{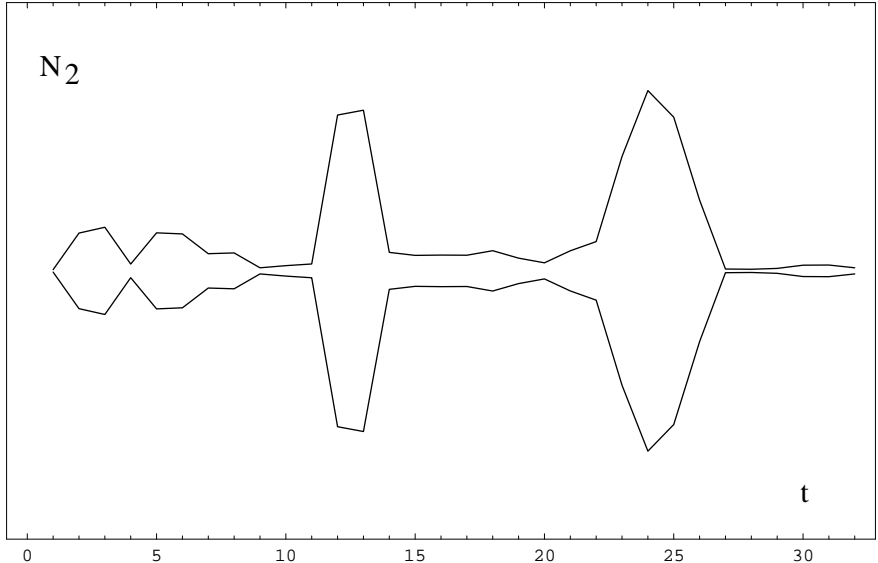}}}}
\vspace{-4cm}
\caption[fig8]{Monte Carlo snapshot of the distribution 
of spatial volumes $N_{2}^{(s+1/2)}(t)$, for $T \equ 
32$, $N_3\equ 16,000$ and $k_0 \equ 6.7$ (that is, 
above the critical value $k_0^c \equ 6.64$). The volumes are
plotted symmetrically about a central horizontal $t$-axis.
}
\label{fig8}
\end{figure}

\begin{figure}[t]
\vspace{-4cm}
\centerline{\scalebox{0.75}{\rotatebox{0}{\includegraphics{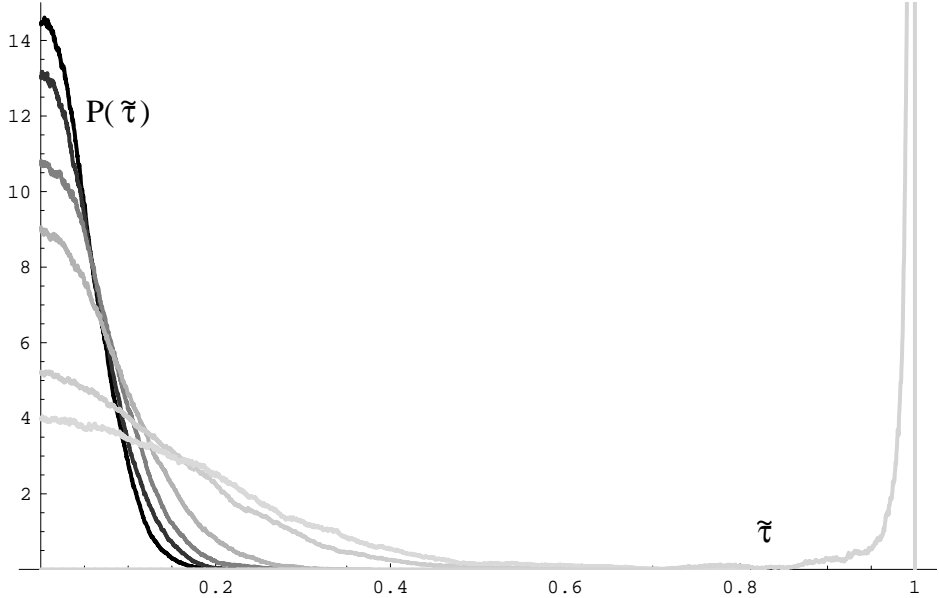}}}}
\vspace{-5cm}
\caption[fig9]{The probability distribution $P(\tilde\tau)$ of
$\tilde{\tau}\equ 
|N_{2}^{(s)}(2)-N_{2}^{(2)}(1)|/N_3$
for $k_0 \equ$ 6.0, 6.1, 6.2, 
6.3, 6.4, 6.42 (highest to lowest peak around $\tilde\tau =0$; 
the value $k_0 =6.42$ corresponds to the phase transition point) 
and 6.44 (distribution peaked around 1); for total volume $N_3\equ
8,000$ and free boundary conditions.}
\label{fig9}
\end{figure}

The physics of this phase can be readily understood. In terms of the
matrix model, a situation with no (2,2)-tetrahedra corresponds
to choosing the coupling constant $\b \equ 0$ in (\ref{matrix}), 
thus reducing the model
to a product of two independent $\phi^3$-matrix models. 
Since a $\phi^3$-matrix model at its critical point describes
two-dimensional Euclidean quantum
gravity, the matrix model analogy strongly suggests that the Lorentzian
3d model for $k_0 > k_0^c$ can be viewed as a product of uncoupled 
2d gravity models. This conclusion seems to be corroborated by our
numerical results. 
Fig.\ \ref{fig8} is a typical ``snapshot'' of a 
space-time geometry, taken during the computer
simulations. The spatial volume $N_{2}^{(s+1/2)}(t)$
is shown as a function of 
the time $t$. Apparently it can change from essentially zero to a
``macroscopic'' size in a single time-step, which implies
that there cannot be any 
correlations between slices separated by a few time-steps.
A different measurement of the correlation between 
successive spatial volumes is depicted in Fig.\ \ref{fig9}.
We have again chosen $T\equ 1$
and free boundary conditions, in order to have spatial slices of
a reasonably large size, but a
qualitatively similar behaviour is expected for $T >1$ too. 
We show the distribution of the 
(normalized) difference 
\beq{ttau}
\tilde{\tau}(1\rightarrow 2) = 
\frac{|N_{2}^{(s)}(t\equ 2)-N_{2}^{(s)}(t\equ 1)|}{N_3}\equiv 
\frac{|N_{2}^{(s)}(2)-N_{2}^{(s)}(1)|}{N_{31}(1)+N_{13}(1)+N_{22}(1)}
\eeq
of the spatial volumes of the initial and final slice. 
For $k_0$ less than the critical $k_0^c\approx 6.42$, $\tilde{\tau}$
is peaked around zero. The peak becomes flatter as $k_0$ approaches its
critical value and immediately beyond $k_0^c$, the distribution approximates
a $\del$-function around $\tilde\tau\equ 1$. 

This last result can be understood as follows. We know from the
simulations that the number of (2,2)-tetrahedra drops to a minimum 
beyond the critical point $k_0^c$. To first approximation, these
tetrahedra therefore do not contribute to the entropy in that region.  
Moreover, since a minimal set of (2,2)-tetrahedra can basically
interpolate between any pair of ``incoming'' (3,1)- and
``outgoing'' (1,3)-configurations, the combinatorics is governed
by the {\it separate} countings of those configurations, subject
only to an overall volume constraint $N_{31}\pl N_{13}\equ N_3\mi N_{22}
\approx N_3\equ const$.
Individually, the configurations at $t\equ 1$ and $t\equ 2$
are simply 2d Euclidean triangulations, whose number for a given
spatial volume $N_{2}^{(s)}$ is known to be proportional to 
$\e^{c N_{2}^{(s)}}(N_{2}^{(s)})^{-5/2}$.
From $\tilde\tau (1\rightarrow 2)
\equ |1-2 N_{13}/N_3|$, and taking into account that the minimal
interpolating $N_{22}$-configuration can be inserted {\it anywhere} in
the incoming and outgoing configurations, one finds 
\begin{equation}
\# (N_{31}\rightarrow N_{13})
\sim \e^{c N_{31}}N_{31}^{-\frac{3}{2}} 
\e^{c (N_3-N_{31})}(N_3-N_{31})^{-\frac{3}{2}} 
\sim (1-\tilde\tau^2)^{-\frac{3}{2}}
\end{equation}
for the combined entropy at fixed volume $N_3$. In agreement
with Fig.\ \ref{fig9}, it shows that the most likely configurations
are those where the entire 3d volume is concentrated at one of the
slices, that is, either $N_{31}\approx 0$, $N_{13}\approx N_3$ or
vice versa.
 
The situation is very different in the phase with $k_0 < k_0^c$. 
Our measurements of $\tilde\tau$ at $T\equ 1$ are 
an indication that also in general in this phase the volumes of
successive spatial slices will be strongly coupled, i.e. their
volume difference will be small. Since this behaviour is not
favoured by the action, the prevalence of such configurations 
must have to do with the presence of the (2,2)-tetrahedra and their
associated entropy (i.e. the number of ways they can combine
with each other and with other tetrahedra
to form interpolating ``sandwiches''). 
This observation turns out to be of great importance, since it
seems to lie at the heart of the emergence of extended geometries
in this phase, which will be described in more detail 
in the next section.
In broad terms, the phase is characterized by $0 < \tau < 1$. 
In principle there 
may be another phase transition at some smaller (possibly negative) 
value $\tilde{k}_0$,  such that 
$\tau \equ 1$ for $k_0 < \tilde{k}_0$ (with no (3,1)- or 
(1,3)-tetrahedra surviving in the continuum limit). 
Indeed, for fixed $N_3$ and negative $k_0$ the action \rf{n2} 
has a minimum for $\tau \approx 1$.
(Configurations with $\tau \to 1$
for $N_3 \to \infty$ can actually be realized.)

Whether or not the system will undergo 
a phase transition for sufficiently small $k_0$ will depend on the
balance between action and entropy, which cannot be determined by
simple qualitative considerations.
We have not investigated this region of the coupling constant space 
further, given the limited
importance of negative gravitational coupling constants 
from a quantum gravity viewpoint, and the fact that 
our computer algorithm is not efficient at small $k_0$.

In summary, we have arrived at the following tentative 
description of the phase diagram of 3d Lorentzian quantum gravity:
the bare inverse gravitational coupling constant has two critical 
values, $\tilde{k}_0^c$
and $k_0^c$ (possibly with $\tilde{k}_0^c\equ\mi \infty$). For 
$k_0 > k_0^c$ the model describes the fluctuations of
an uncorrelated set of two-dimensional spatial geometries and has nothing to 
do with a three-dimensional theory of gravity. 
Also for $k_0 < \tilde{k}_0^c$ the space-time geometry degenerates, since
the spatial slices at integer $t$ completely disappear from the theory.
These two ``extreme'' regions of the phase diagram can be regarded
as artifacts of our particular way of setting up the discretized theory.
They may  
be seen as remnants of the phases of degenerate geometries observed
previously in 3d Euclidean quantum gravity \cite{former}. 
However, unlike the
Euclidean theory, Lorentzian gravity possesses a large region 
$\tilde{k}_0^c < k_0 < k_0^c$ of coupling constant space 
where the quantum geometry is extended and well-behaved, and 
whose description we shall turn to next.

\subsection{The phase of extended geometry} 

\begin{figure}[t]
\vspace{-4cm}
\centerline{\scalebox{0.75}{\rotatebox{0}{\includegraphics{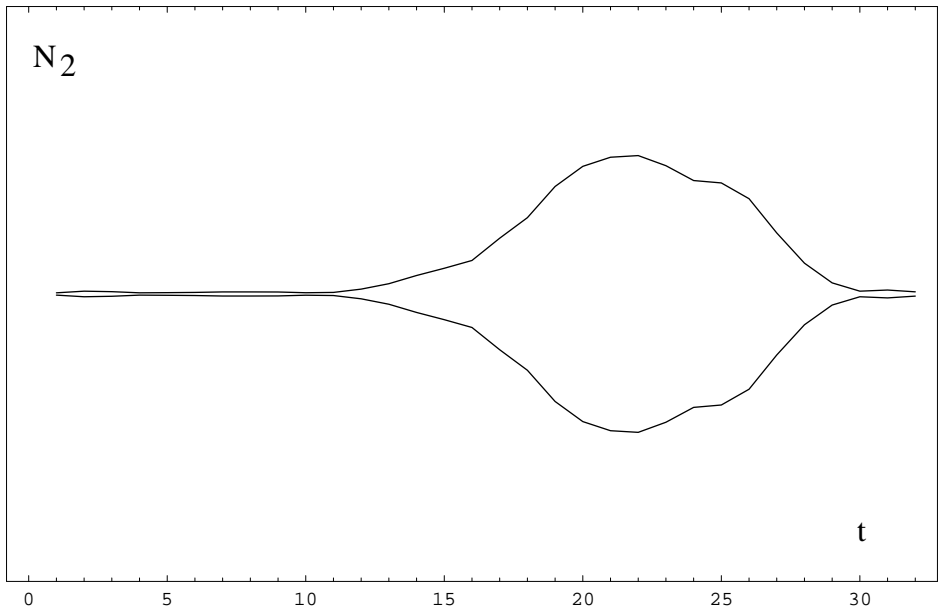}}}}
\vspace{-5cm}
\caption[fig10]{Monte Carlo snapshot of the distribution $N_{2}^{(s+1/2)}(t)$
of spatial volumes, for $T \equ 
32$, $N_3\equ 16,000$ and $k_0 \equ 5.0$ (that is, below the critical value 
$k_0^c \equ 6.64$).
The volumes are
plotted symmetrically about a central horizontal $t$-axis.}
\label{fig10}
\end{figure}

Let us now analyze the structure of the phase of intermediate gravitational
coupling, $\tilde{k}_0^c < k_0 < k_0^c$, where all types of tetrahedral
building blocks contribute non-trivially. 
Quite remarkably, and unlike in the phase where $k_0 > k_0^c$ 
we observe here the emergence of well-defined
three-dimensional configurations.
Fig.\ \ref{fig10} shows a snapshot of a typical geometry at $k_0\equ 5.0$,
consisting of 16,000 tetrahedra, for $T\equ 32$. 
(As in the previous Fig.\ \ref{fig8}, we plot -- symmetrically around an
arbitrary axis -- the spatial volume $N_{2}^{(s+1/2)}(t)$ as a function of 
$t$.)
Following the computer-time history 
of this extended object, it is clear that although it does indeed
fluctuate, the fluctuations take place around a three-dimensional 
object of well-defined linear extension.\footnote{
A trivial mode of fluctuations are the translations in time-direction. Due 
to the periodicity of the boundary condition the ``centre of volume'' 
of the 
extended configuration performs a random walk in the $t$-direction.} 
{\it The emergence of a ground state of extended quantum 
geometry is a highly non-trivial property of the Lorentzian model,
since we have at no stage put in a preferred background geometry by hand.}
No structures of this kind have ever been observed in
dynamically triangulated models of {\it Euclidean} quantum gravity.
It underscores the fact that the Lorentzian models are genuinely
different and affirms our conjecture \cite{ajl} that in $d\geq 3$ they are
less pathological than their Euclidean counterparts.

For a fixed $k_0$ and $N_3$, an overall ``spherical'' shape as 
shown in Fig.\ \ref{fig10} is found only for sufficiently large $T$.
At small $T$, one observes 
a uniform distribution of spatial volumes $N_{2}^{(s+1/2)}(t)$ as a 
function of 
$t$. As $T$ increases, the bulk of the space-time volume ``condenses'' 
into a region with a well-defined extent $T_u <T$ in time-direction
(which we will call the {\it universe}),
leaving over a thin stalk of minimal spatial radius everywhere else
along the $t$-axis. We will from now on choose $T$ sufficiently large,
so that $T>T_u$ for all volumes under consideration. 
We are interested in the ``cosmological'' properties of this extended
universe, i.e. its geometric properties at large scales. 
Our data for the scaling of the time extent $T_u$ of the universe as a
function of the total volume are consistent with 
\beq{n3}
T_u \sim N_3^{1/3}.
\eeq
(We discuss below how a quantitative measure of $T_u$ is obtained.)
Similarly, by an {\it independent} measurement of the volumes 
$N_{2}^{(s+1/2)}(t)$
of spatial slices that lie within the universe, 
we have found a scaling behaviour consistent with
\beq{n4}
\la N_{2}^{(s+1/2)}(t)\ra \sim N_3^{2/3}.
\eeq
The relations \rf{n3} and \rf{n4} support an interpretation of the 
universe as a genuinely three-dimensional object. 
There is of course no a priori reason that a ground state in a 
non-perturbative theory of quantum gravity
(if it exists) should
bear any resemblance with a (semi-)classical geometry. 
Let us explain briefly how such geometries (and, more specifically,
classical solutions) {\it might} still make an appearance in
this context.

For the simplest compact space-time topology, the solution to the 
{\it classical} Einstein equations with 
Euclidean signature and
a positive cosmological constant $\L$ is the round three-sphere
(of constant positive scalar curvature) 
with radius $R_{S^3} \sim \L^{\mi 1/2}$. 
Solving the equations 
with the constraint of fixed space-time volume $V$ is equivalent to 
introducing an effective cosmological constant 
$\L_{eff} \sim V^{ \mi 2/3}$.
We are not aware of a classical solution with positive $\Lambda$ and
topology $S^1\times S^2$ (which is the topology used in our
simulations).
However, for our purposes we can ``adapt'' the $S^3$-solution to this
topology by cutting away two small open balls at two opposite points 
of an $S^3$-configuration with radius
$R_{S^3} \sim \L_{eff}^{\mi 1/2}\sim V^{1/3}$ and attaching 
a thin cylinder $I\times S^2$ (with spatial radius at the cut-off
scale) to the holes. This will produce a geometry of the kind
shown in Fig.\ \ref{fig10}.
Although it is not strictly speaking a solution 
to Einstein's equations, it is ``almost as good'' from the point of
view of the path integral, since -- 
independent of its metric properties --
the contribution of the stalk 
to the action is negligible (because it does not grow proportionally
to the three-volume).

Suppose for the moment that the round $S^3$-solution 
corresponded to a (local) minimum of the action. 
Then the singular
``solution'' of topology $S^2\times S^1$ constructed above would also 
represent a (local) minimum of the action, and would therefore be 
as relevant as the $S^3$-solution {\it in the quantum theory}. 
Unfortunately, the argument is not quite as simple, because 
the classical continuum Einstein action is unbounded from below, 
due to the presence of a kinetic term of the ``wrong'' sign,
coming from the conformal mode of the metric.
However, since the conformal mode is not a propagating degree of
freedom in either classical general relativity or
in canonical quantizations, it should not cause any problems in
a correct, non-perturbative path-integral quantization of gravity, 
not even in the Euclidean sector\footnote{The continuum path integral 
in proper-time gauge is discussed in \cite{dl}.}. 
In such a quantum 
theory the {\it effective} action should be bounded from below
and semi-classical saddle-point considerations of the kind made
above may again be appropriate.

\begin{figure}[t]
\vspace{-4cm}
\centerline{\scalebox{0.75}{\rotatebox{0}{\includegraphics{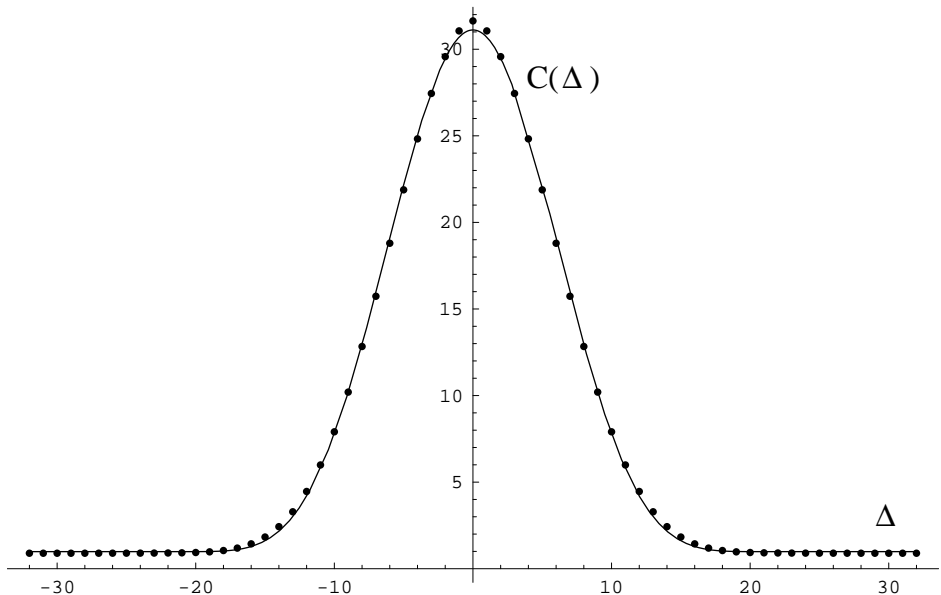}}}}
\vspace{-5cm}
\caption[fig11]{The correlator $C(\Del)$ with 
$T\equ 64$ and $N_{3}\equ 32,000$. Dots are the measured 
values (error bars less than dots), and the curve is fitted from 
the sphere solution described in the text.}
\label{fig11}
\end{figure}

We have measured the correlation function   
\beq{n5}
C(\Del) = \frac{1}{{T}^2}\sum_{t=1}^{T}\la 
N_{2}^{(s+1/2)}(t) N_{2}^{(s+1/2)}(t+\Del )\ra
\eeq 
as a function of the displacement $\Del$
to determine the scaling of $T_u$ with the
space-time volume $N_3$.
This correlator has the advantage of 
being translation-invariant in $t$ and allows for a precise 
measurement by averaging over many independent configurations. 
From the typical shape of the space-time configurations 
we expect $C(\Del)$ to be of the order of the spatial cut-off 
if $\Del > 2 T_u$. Fig.\ \ref{fig11} illustrates the result of  
our measurements of $C(\Del)$, with the dots representing the
measured values.
The theoretical curve to which we are fitting corresponds to the 
``fake sphere'' described above, with the radius of the $S^{3}$
and the spatial radius of the thin cylinder attached to it 
as free parameters. 
For this ``spherical'' geometry we then perform the integral (the sum) 
in \rf{n5}, without the average $\la \cdot \ra$.
As is evident from Fig.\ \ref{fig11}, the volume distribution
associated with this fixed geometry gives a rather good fit to
our data. This provides some evidence that we can 
ignore the quantum average implied by $\la \cdot \ra$, and that 
our universes behave
semi-classically, {\it at least
as far as their macroscopic geometric properties are concerned}.
We should mention that our ``$S^3$-solution'' is not
singled out uniquely, since the choice of a Gaussian shape in the 
$t$-direction gives a fit of comparable quality. 

For various space-time volumes $N_{3}$ (typically 8, 16, 32 and 64k)
we have determined the radius $R_{S^3}$ of $S^3$ 
from the fits to the measured 
$C(\Del)$. From this, we have finally found $\a = 0.34\pm 0.02$
as the best exponent in the scaling relation
\beq{n6}
R_{S^3}(N_{3}) = N_{3}^\a.
\eeq
The same value is obtained using other ways to extract $T_u$, 
lending additional support to the three-dimensional
nature of our universes.

%\vspace{12pt}
We will now take a closer look 
at the geometry of the two-dimensional spatial slices. 
If they could be described as typical triangulated surfaces in 
2d Euclidean quantum gravity, they would not behave like smooth
2d geometries (when described in terms of geodesic distances),
but rather like fractal spaces with Hausdorff dimension 
$d_{H}^{\rm sp}\equ 4$ \cite{d4haus1,d4haus2}.
By contrast, typical {\it space-time} surfaces contributing to the
path integral of 2d Lorentzian quantum gravity can be viewed as
two-dimensional, as shown in \cite{al,aal1,aal2}.  

The spatial slices at constant integer $t$ are obviously Euclidean 
in nature, but it is not
immediately clear how they will behave, since they appear as part
of a larger foliated space-time geometry, and are coupled to each
other in a non-trivial way. We have tried to extract the Hausdorff
dimension $d_{H}^{\rm sp}$ of the spatial slices lying {\it inside} 
the spherical
universe, using the geodesic distance inherited
from the 3d geometry, and employing 
techniques developed in the context of 2d dynamically triangulated 
Euclidean quantum gravity \cite{numhaus1,numhaus2}.
Unfortunately, the quality of our measurements is not very
satisfactory, since the spatial volumes $N_{2}^{(s)}(t)$ 
of the individual
slices are rather small (typically of the order of 1k).

One can obtain better data by using simulations with small $T$
(so that $T <T_{u}$ and no universe can form),
but it is not entirely clear whether this will leave the spatial 
Hausdorff dimension unchanged.
Our measurements for small $T$ point to a value around
$d_H^{\rm sp} \equ 3.4\pm 0.4$ (the measurements for larger $T$
are compatible with this value, but their error bars are 
considerably larger). 
If our experience with the 2d Euclidean gravity 
simulations is anything to go by, this probably implies 
$d_H^{\rm sp} \equ 4$, but so far this has to remain merely a conjecture. 
At any rate, these somewhat preliminary results highlight
that the detailed, microscopic geometry of the universe may be
rather complicated, although its macroscopic properties resemble
that of a semi-classical object.

Attempts to measure the Hausdorff dimension $d_{H}$ 
of the entire space-time
(as opposed to that of individual spatial slices) 
have not yet led to unambiguous results. One wants to confine the
measurement to the spherical universe, where again one runs into
difficulties because of its relatively small radius. In addition,
one needs a dynamical definition of where the universe begins and
ends (along the $t$-direction), and must make sure that the result
is independent of the particular prescription adopted.
From the limited data collected (using 
the geodesic link or dual link distance, 
in the sense in which this notion is usually defined in dynamical 
triangulations) we conclude that
the Hausdorff dimension is most likely larger than three. 

%\vspace{12pt}

Another important result concerns 
the relation between the geometries of different $k_0$, in the phase
where $k_0 <k_0^c$. In the numerical simulations we have observed
the following: 
\begin{itemize}
\item[(i)] the distributions as functions of 
$t$ can be made to coincide for different $k_0$ 
by rescaling the time, $t \to f_{\rm ti}(k_0) t$ or alternatively
$a_t \to f_{\rm ti}(k_0) a_t$, where $a_t$ is the link length in time 
direction. 
This is illustrated by the $N_{2}^{(s+1/2)}$-$N_{2}^{(s+1/2)}$ 
correlator $C(\Del)$, Fig.\ \ref{fig12}, where
we show both the actual and the rescaled distributions.
\begin{figure}[t]
\vspace{-4cm}
\centerline{\scalebox{0.5}{\rotatebox{0}{\includegraphics{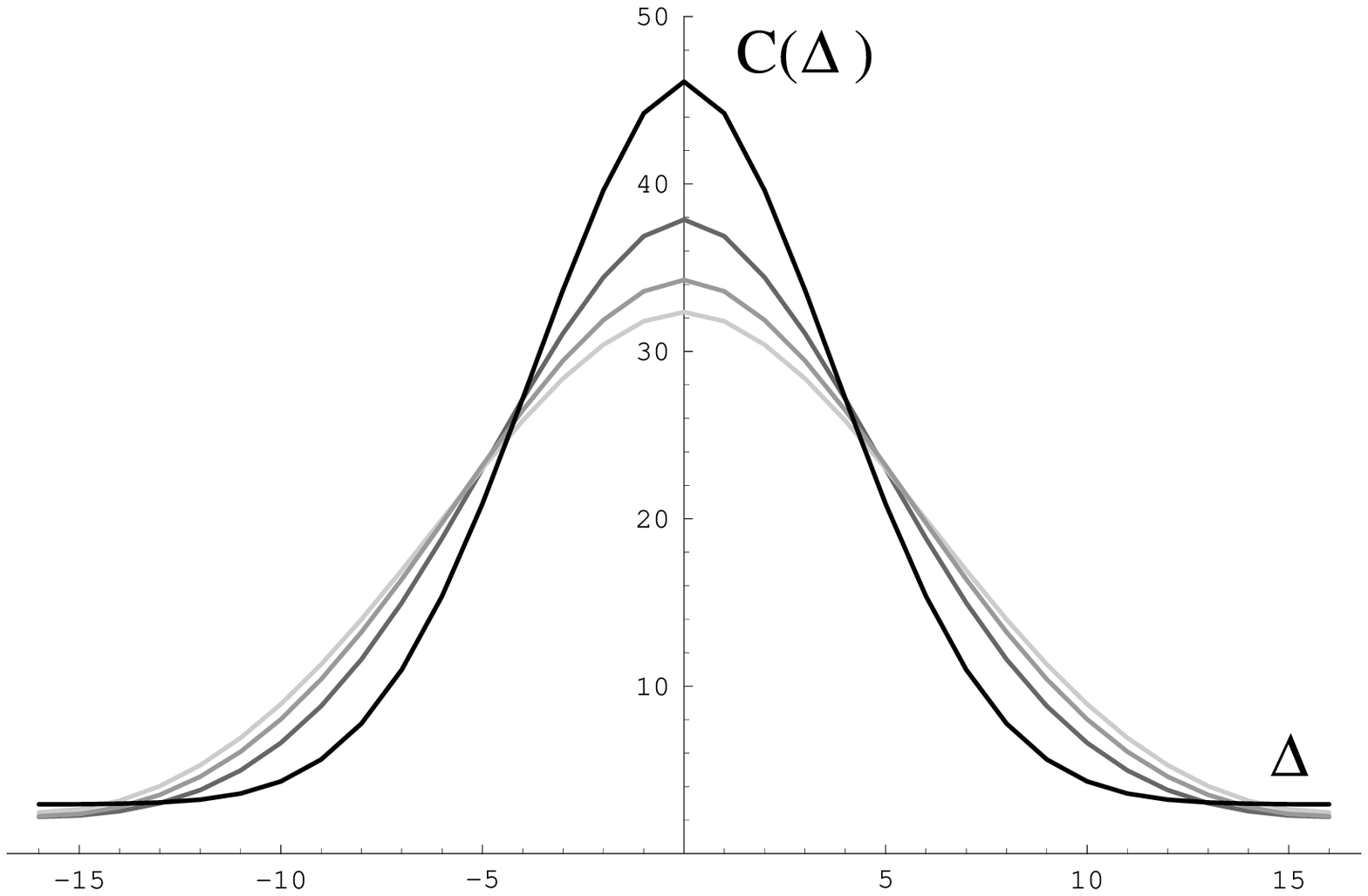}}}}
\vspace{-6cm}
\centerline{\scalebox{0.5}{\rotatebox{0}{\includegraphics{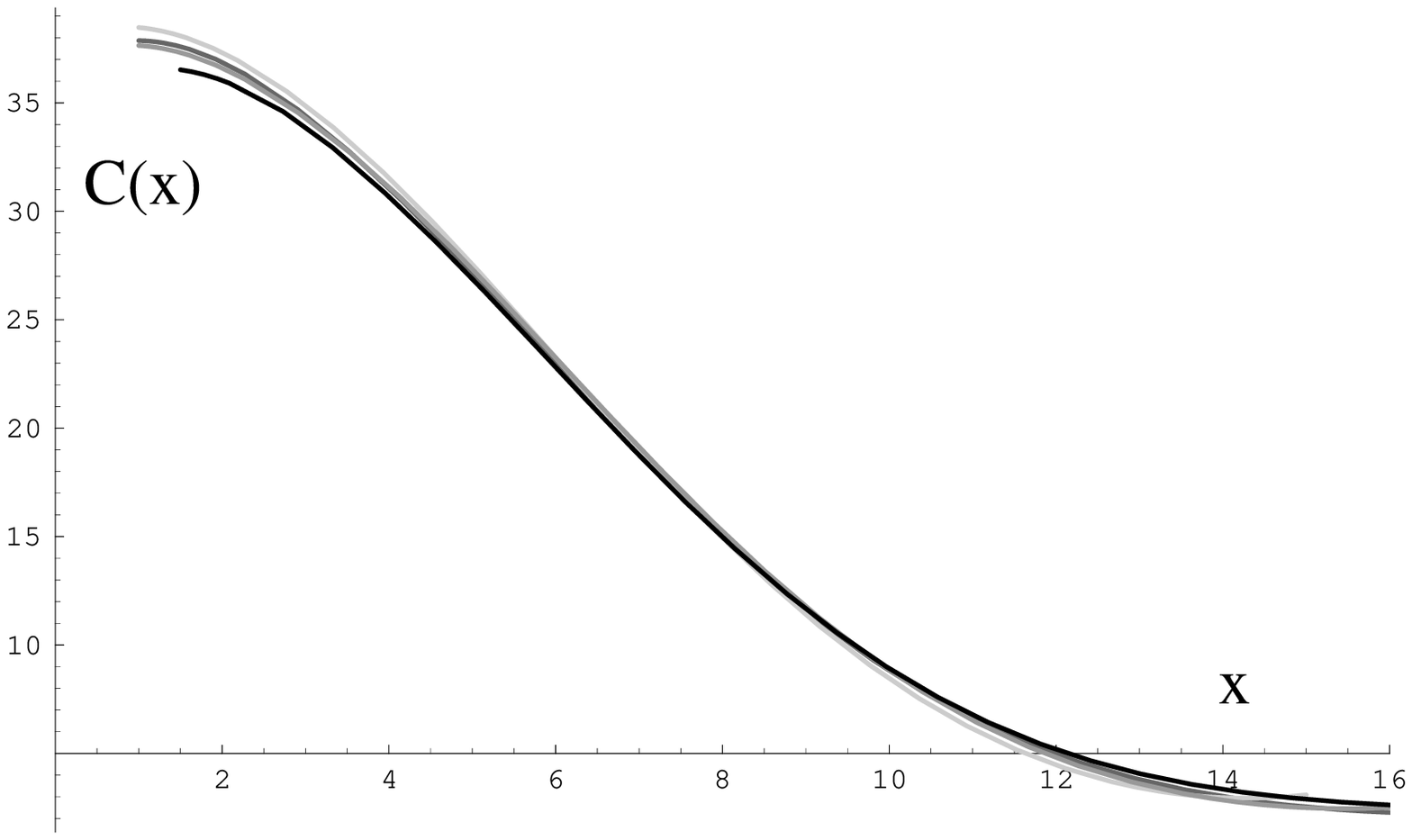}}}}
\vspace{-3cm}
\caption[fig12]{The correlator $C(\Del)$, eq.\ \rf{n5} with 
$T\equ 32$ and $N_3\equ 16,000$ measured for $k_0\equ 3.0$, 4.0,
5.0 and 6.0 (lowest to highest peaks). The upper figure shows
the actual distribution, the lower one the rescaled distributions
for positive $\Delta$ (it is symmetric in $\Delta$).
The variable $x$ is a
rescaled version of the time distance $\Delta$.}
\label{fig12}
\end{figure}
\item[(ii)] The distributions measured in the spatial slices 
from {\it inside} the universe can be made to 
coincide for different $k_0$ by rescaling the 
spatial link distance $a_s \to f_{\rm sp}( k_0) a_s$, where $a_s$ is 
the length of the spatial links. 
This is illustrated in Fig.\ \ref{fig13} for the distributions 
of 2d volumes $S(l)$ of {\it spatial} spherical shells of 
(link) radius $l$, measured for various values of $k_0$. 
(The shell volume $S(l)$ is obtained by counting the number of
vertices separated from a given vertex $v$ by a minimal 
link distance $l$. Note that this spherical shell is precisely 
what is measured to determine the Hausdorff dimension $d_H^{\rm sp}$ 
of the spatial slices.)
\begin{figure}[t]
\vspace{-4cm}
\centerline{\scalebox{0.75}{\rotatebox{0}{\includegraphics{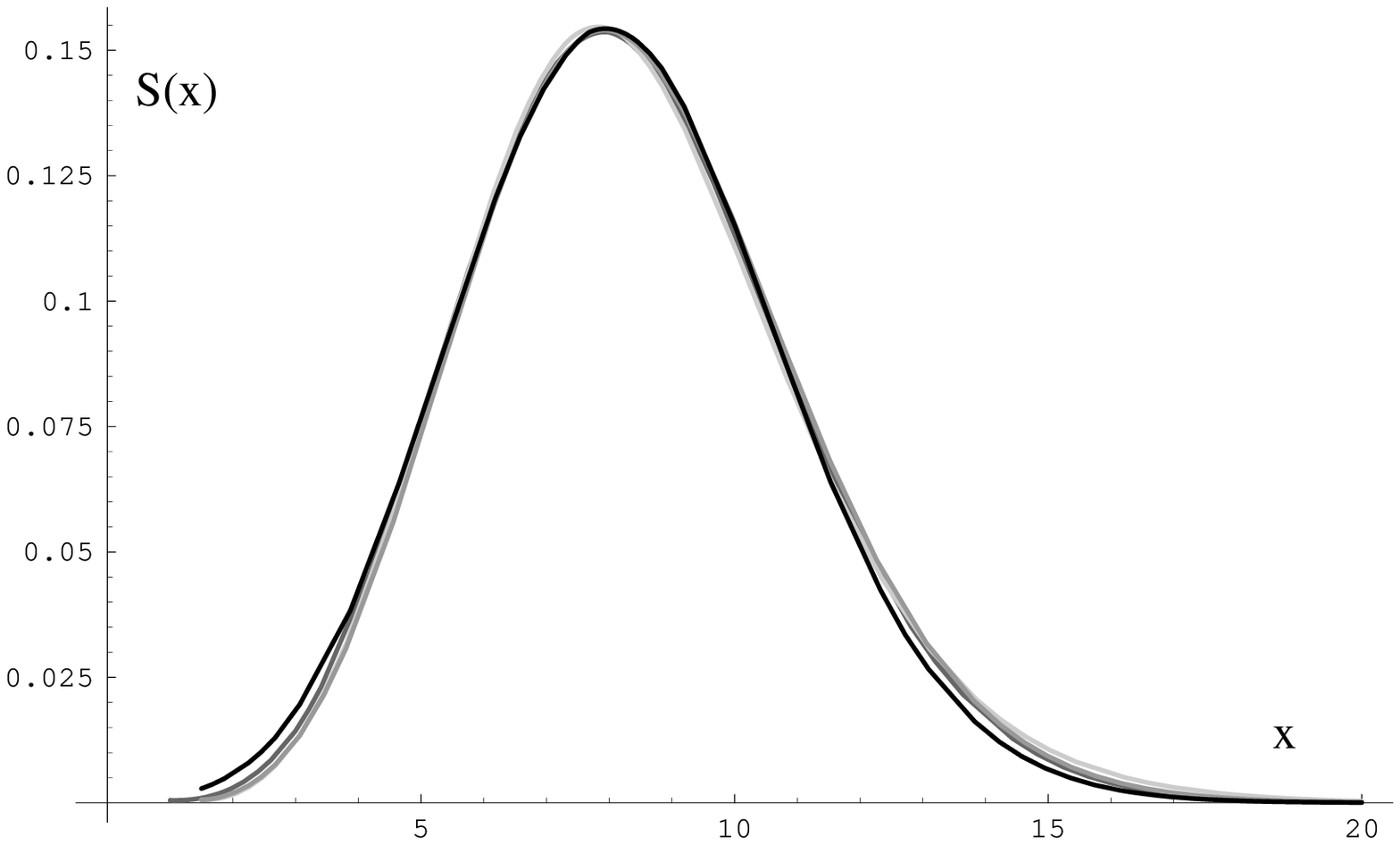}}}}
\vspace{-5cm}
\caption[fig13]{The 2d volume $S(x)$ of spatial spherical shells,
measured only on slices inside the spherical universe,
for various values of $k_0$ and rescaled. The variable $x$ is a
rescaled version of the radius $l$.}
\label{fig13}
\end{figure} 
\item[(iii)] Within the numerical accuracy we find that
$f_{\rm ti}(k_0)=f_{\rm sp}(k_0)$. In fact, the rescaling of the correlator
$C(\Del)$ (Fig.\ \ref{fig12}) was obtained 
by simply using the values $f_{\rm sp}(k_0)$ (see Table\ 1)
determined from the fit $S(l)$ (rather than
by finding the best value for $f_{\rm ti}(k_0)$).
\end{itemize}  

\begin{table}[h]
\begin{center}
\begin{tabular}{||c||c||}
\hline
$k_0$ & $f_{\rm sp}(k_0)$ \\
\hline
3.0 &$ 0.84 \pm .02$ \\
\hline
4.0 & $0.91 \pm .02$ \\
\hline
5.0 & 1.00          \\
\hline
6.0 & $1.23 \pm .03$ \\
\hline
\end{tabular}
\end{center}
\caption{The spatial scaling factor $f_{\rm sp}(k_0)$, extracted
from the distributions $S(l)$ for various values of $k_0$.}
\end{table}
On the basis of these correlator measurements we conjecture that
{\it the value of
the bare inverse gravitational coupling $k_0\in ]\tilde{k}_0^c,k_0^c[$ 
merely sets the overall length scale of the universe, and otherwise
does not affect the physics of the model}.

We should point out that the average total integrated
curvature is not independent of $k_{0}$. Subtracting the cosmological
term from the action \rf{app10}, one finds
\beq{curv}
\int d^{3}x \sqrt{\det g}  R \longrightarrow \pi a N_{3}
(\tau (k_{0}) +(12 \kappa -5) -8 \frac{T}{N_{3}})\sim
\tau (k_{0}) -0.298 -8 \frac{T}{N_{3}},
\eeq
where the parameter $\tau$ (defined in eq.\ \rf{tau}) now depends
dynamically on $k_{0}$ through the ensemble average. Comparing 
with our measured curve for $\tau$ in Fig.\ \ref{fig6}, one can read 
off that the total curvature vanishes around $k_{0}\approx 5.0$.
For smaller $k_{0}$, it becomes negative and for larger $k_{0}$
positive. Nevertheless, in line with our conjecture above 
we expect the curvature-curvature 
correlators to follow the pattern of the already measured 
correlators (i.e. to observe a simple $k_{0}$-dependent scaling
behaviour), but this remains to be verified.

\section{Summary and discussion}

\begin{figure}[ht]
%\vspace{-.6cm}
\centerline{\scalebox{0.6}{\rotatebox{0}
{\includegraphics{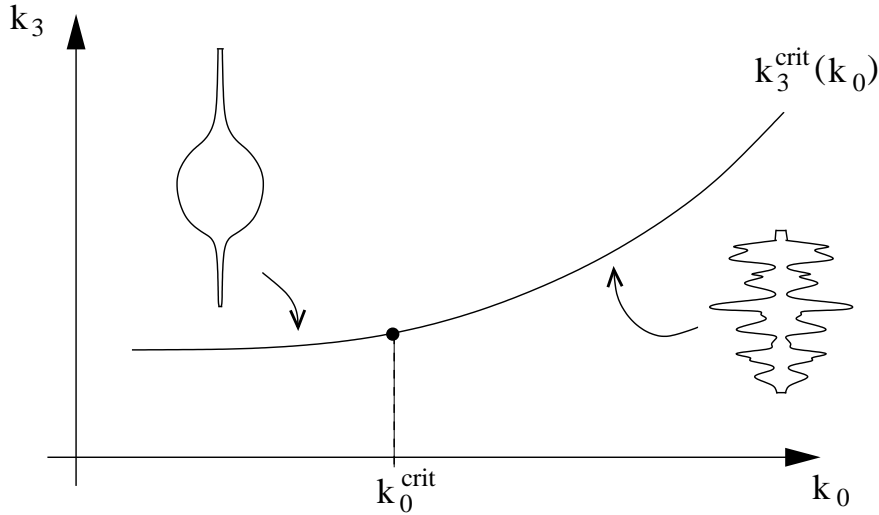}}}}
%\vspace{-.6cm}
\caption[phaselor]{The phase diagram of Lorentzian dynamical
triangulations in three dimensions.}
\label{phaselor}
\end{figure}

\begin{figure}[ht]
%\vspace{-.6cm}
\centerline{\scalebox{0.6}{\rotatebox{0}
{\includegraphics{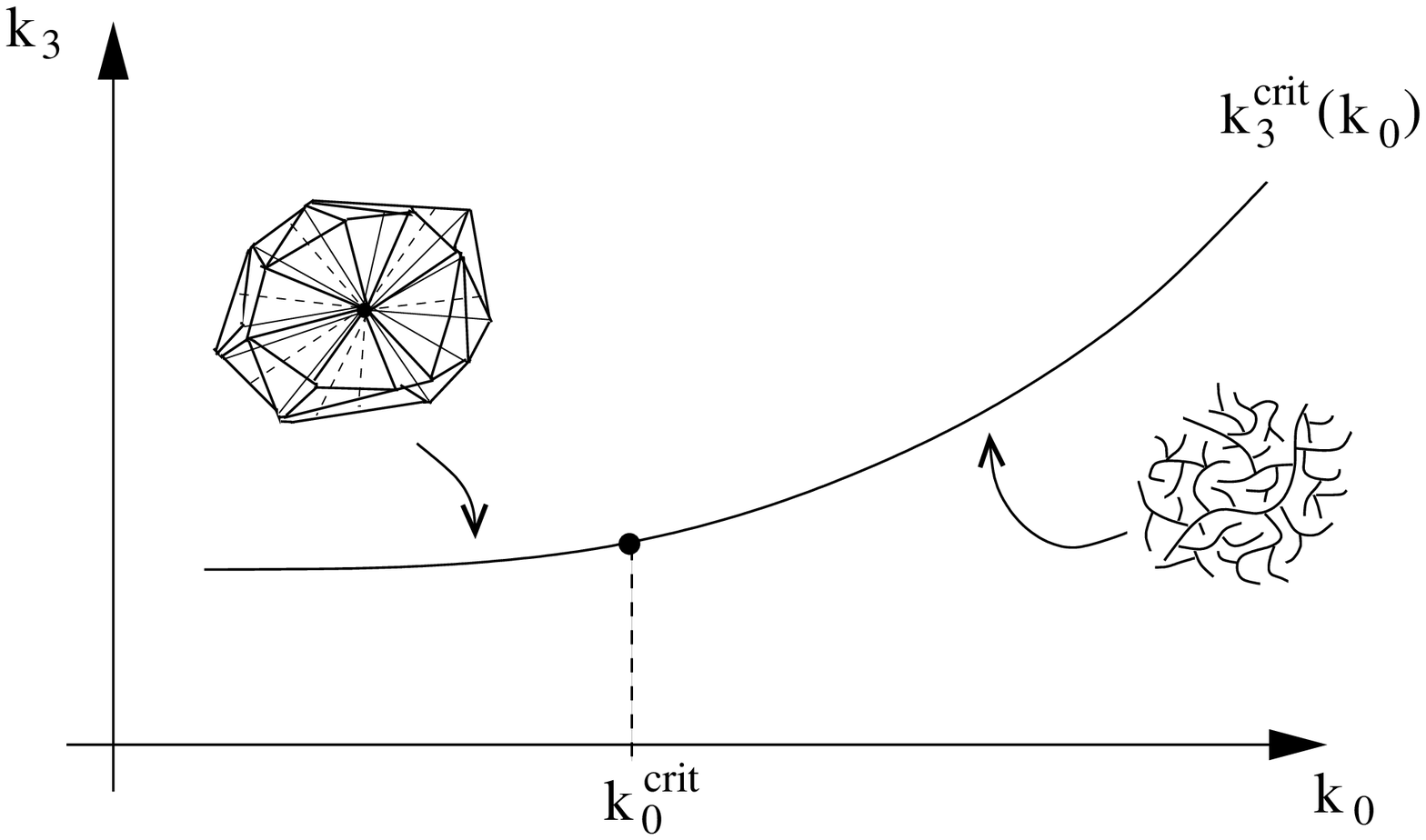}}}}
%\vspace{-.6cm}
\caption[phaseeu]{The phase diagram of Euclidean dynamical 
triangulations in three dimensions.}
\label{phaseeu}
\end{figure}

In this paper, we have analyzed the phase structure 
of the discretized model of three-dimensional Lorentzian
gravity defined in \cite{ajl} with the help of computer simulations.
The phase diagram, Fig.\ \ref{phaselor}, should be compared with that 
of the Euclidean theory, depicted in Fig.\ \ref{phaseeu}. 
Although the overall phase structure is similar, with a first-order 
transition at some intermediate value $k_{0}^{c}$, the quantum
geometries of the phases on either side of the transition are
very different in both cases, as indicated in the drawings.

In the Euclidean case \cite{former}, one finds
a ``crumpled'' phase at small $k_{0}$, dominated by
configurations of very large Hausdorff dimension $d\approx\infty$
(these are simplicial manifolds where roughly speaking 
any two vertices are a 
minimal distance apart). Above the first-order
transition at $k_0^c$, the system is in a branched-polymer
phase of highly branched geometries
(with a fractal dimension $d_H=2$). 
Unfortunately, neither of these phases seems to have
a ground state that resembles an extended geometry of dimension
$d\geq 3$. 

Another approach to {\it Euclidean} gravity was advocated in 
\cite{hamwil} within the quantum Regge calculus program.
The phase structure found in the numerical simulations of this 
model resembles those of Figs.\ \ref{phaselor} and
\ref{phaseeu} at least superficially in exhibiting a ``rough phase''
for small and a ``smooth phase'' for large Newton's constant.
However, this model is inequivalent to the dynamically 
triangulated models we have been discussing, since 
in the Regge approach these two phases seem to be separated by a
second-order phase transition and associated divergent
curvature fluctuations, indicating the
presence of propagating field degrees of freedom (cf. our general
discussion in Sec.2).
How this can be related to the topological character of 3d
quantum gravity manifest in canonical treatments of the
theory is unclear. 
 
The situation in Lorentzian dynamically triangulated 
gravity is completely different.
Although we find a weak ``remnant'' of the Euclidean degeneracy
for $k_{0} >k_{0}^{c}$, where space-time decouples into a
sequence of uncorrelated two-dimensional slices, there is
a whole phase below $k_{0}^{c}$ where the geometry is extended,
with macroscopic scaling properties characteric of a 
three-dimensional universe. 
Quite remarkably, this is an example in three dimensions
of the emergence 
of a well-defined ground state of geometry in a non-perturbative
state sum for gravity. 
In a continuum language, this is the ground state of an {\it effective} 
action, where entropy contributions (in other words, the {\it measure})
play a crucial role. Apparently in our model these contributions are
such that they outbalance potential conformal divergences coming
from the Wick-rotated action (otherwise a well-defined ground state
could not exist).
From the evidence gathered so far, the physics in this 
extended phase is {\it independent} of the precise value of the 
bare gravitational coupling $k_0$. In the correlation functions we 
have measured, $k_0$ merely serves to set an effective overall
length scale. 

As argued in Sec.\ 2, these findings strongly favour
a situation where the gravitational coupling is not renormalized, and
no fine-tuning of $k_{0}$ is needed to approach the 
continuum limit. 
This limit coincides automatically with the infinite 
(lattice) volume limit, which we obtained by fine-tuning the 
cosmological coupling constant. 
Continuum physics is then extracted by taking the limit as
$N_{3}\rightarrow\infty$ and $a\rightarrow 0$, while 
keeping the three-volume $V_{\rm cont} := N_3 a^3$
constant. In this setting, no genuine field degree of freedom is
present since there is no divergent correlation length 
associated with fine-tuning $k_0$ to the critical point 
of a second-order phase transition. 

As a result of our investigations, we have good reasons to believe
that 3d Lo\-rentz\-ian quantum gravity, as defined through our discrete,
dynamically triangulated model, exists as a continuum theory.
Since so far this theory is not given in an explicit
analytical form, the question arises of how to make contact with
already existing quantizations of three-dimensional gravity. 

At least in spirit, our formulation is related to canonical
approaches using the trace of the extrinsic curvature as a
time variable, the so-called ``York time'', with a 
conjugate Hamiltonian determining the time evolution of the
system \cite{moncrief,carlip}. 
However, this approach only works for genus $g\geq 1$, and 
the only case where the 
canonical quantum theory and the Hamiltonian operator are 
reasonably explicit is $g\equ 1$, where the spatial
slices have torus topology. We are not aware of any quantum
observables that have been calculated in the case of spherical
slices, and which we could try to compare to. (For $S^2$-slices,
there are no non-contractible holonomies and the reduced
phase space is zero-dimensional.)
We could in principle repeat the simulations for toroidal
spatial slices, although the finite-size effects will be larger for
this more complicated topology (and for $T\equ 32$, 64, 
our spatial slices are rather small). 

Similarly, it is in principle straightforward to enlarge 
the Lorentzian model to include coupling to matter fields.
This has already been done in two-dimensional 
Lorentzian quantum gravity, with a clear 
motivation in mind, namely, to understand the status of the
$c \equ 1$ barrier in general 2d gravity models. 
We showed that this disease of 2d Euclidean quantum gravity 
can be avoided by working with Lorentzian geometries \cite{aal2}.

It would provide a strong incentive for considering either of these
generalizations if there were definite predictions from
continuum formulations of 3d quantum gravity with and without
matter for observables measurable in the
computer simulations (for example, correlation functions of
the type we have been studying).

A next important step in our analysis of 3d Lorentzian
quantum gravity will be the derivation of the
explicit form of the Hamiltonian in the continuum limit.
We can in principle obtain the matrix elements of the
transfer matrix $\hat T$ between two successive triangulated
two-geometries $g_i$, as the solution to a combinatorial
problem \cite{toappear}. Expanding the matrix elements according
to 
\beq{con2}
\langle g_2 |\hat T |g_1\rangle =
\la g_2|\; \e^{-a \hat{H}}\;|g_1\ra = 
\la g_2| \;\big( 1 - a \hat{H}+ O(a^2)\big)\;|g_1\ra,
\eeq
one can extract the Hamiltonian operator $\hat{H}$. 
A similar approach was successful in 2d Lorentzian gravity,
where the regularized transfer matrix could be calculated,
and its continuum limit taken in a straightforward 
way\footnote{The formula for the Hamiltonian in
\cite{al} contains a typo; see \cite{rew2d} for the correct 
expression.}. The resulting Hamiltonian agreed with the 
one obtained by continuum formal manipulations  
in proper-time gauge \cite{nakayama}, showing that the educated 
guesses made in this paper were justified. 

This calculation can  
be generalized to our 3d Lorentzian gravity model, but the
matrix-model methods will probably only work 
in the case of a spherical spatial topology. 
A direct comparison with canonical quantum gravity would then 
require a canonical continuum quantization 
in proper-time gauge, with spatial topology $S^2$.
 
Let us conclude by pointing out an interesting consequence of 
our arguments
that would follow if the second critical point $\tilde{k}_0^c$ 
(whose value we did not measure)
was {\it negative}. In this case,
the theory with bare coupling constant $k_0 \equ 0$ would
lie inside the extended phase. This implies that we could
start from a discretized gravity action {\it with
the cosmological term alone}, and still obtain the same
continuum theory. This may seem a radical suggestion, because
the classical theory of the action 
\beq{con3}
S = \Lambda \int \d^3 x\;\sqrt{\det g(x)}
\eeq
is trivial (it does not contain any time-derivatives). 
However, there is no logical contradiction, since further
non-trivial contributions to the (effective) action can be
generated through the non-perturbative evaluation of the path
integral. 
In fact, 2d Euclidean quantum gravity provides 
a good illustration of this mechanism. 
There the action is given by 
eq.\ \rf{con3}, but substituting $\d^3 x\to \d^2 x$. Nevertheless,
the effective quantum theory in conformal gauge is 
described by the highly non-trivial quantum Liouville theory.

What have we learned from our exploration about our ultimate goal,
the construction of quantum gravity in four dimensions?
We invented the discrete {\it Lorentz\-ian} models in the hope
that they may lead to a better description of physical
four-dimensional space-time, which after all has a 
Lorentzian signature.
We also conjectured in \cite{ajl} that in the continuum limit 
the causality constraints
imposed on each geometry in the state sum 
may lead to a suppression of the degenerate phases of highly
fractal geometry found in the Euclidean models for $d\geq 3$.
From the evidence presented in this work, this is indeed what
happens in three dimensions. Moreover, we saw the emergence
of a ground state of extended three-dimensional geometry
in the Lorentzian case.
As already observed in d=2, also in three dimensions the
Euclidean and Lorentzian models correspond to completely
different continuum theories, re-iterating that these two
``sectors'' of the gravitational quantum theory are
not related by a simple analytic continuation in time \cite{al,ackl}.

We are very encouraged by these results, since they indicate
that also in d=4 completely different geometries will
dominate the Wick-rotated path integral, compared with the
Euclidean theory. The physics that the four-dimensional model 
should describe,
if it were to lead to a non-perturbative theory of quantum gravity,
must of course be very different from that found in two and
three dimensions. In particular, the critical behaviour of the 
regularized theory should reflect the presence of physical, propagating 
field degrees of freedom. 
In the context of the statistical models we are considering,
the simplest realization would be in terms of a second-order 
phase transition. 
This possibility is apparently not realized
in the dynamically triangulated {\it Euclidean} gravity models. 
However, there is by now plenty of evidence that the Lorentzian model for 
quantum gravity defined in \cite{ajl,ajl2} is sufficiently different 
to make it
a new, promising candidate for a non-trivial non-perturbative theory 
of quantum gravity in four dimensions.

\subsection*{Acknowledgements} All authors
acknowledge support by the
EU network on ``Discrete Random Geometry'', grant HPRN-CT-1999-00161, 
and by ESF network no.82 on ``Geometry and Disorder''.
In addition, J.A. and J.J. were supported by ``MaPhySto'', 
the Center of Mathematical Physics 
and Stochastics, financed by the 
National Danish Research Foundation, and
J.J. by KBN grants 2P03B 019\,17 and 998\,14.

\section*{Appendix 1}

In this appendix we collect some formulas for dynamically triangulated
three-geo\-me\-tries, which were used in deriving various
forms of the discrete Einstein action in the main text. We will work in
the Euclidean sector of the theory, and for simplicity choose all
tetrahedra to be equilateral (that is, $\alpha\equ\mi 1$ and
$l_{\rm space}=l_{\rm time}=a >0$). 
The curvature of a 3d piecewise linear manifold is concentrated 
at its links. The contribution to the total curvature associated
with each link $l$ is given by the link length $a$ times 
the deficit angle 
$$
\delta_{l}=2\pi -\sum_{\sigma_{i}\supset l}\theta_{i},
$$ 
where the sum is
taken over all tetrahedra $\sigma_{i}$, $i=1,\dots, o(l)$, 
sharing the link $l$,
and $\theta_{i}$ is the dihedral angle associated with the $i$'th
tetrahedron. For an equilateral three-complex, all dihedral
angles are identical, 
\beq{app1}
\th = \arccos\frac{1}{3} \equiv \k \pi,
\eeq
and the curvature term of the Einstein action becomes
\beq{app3}
\frac{1}{2}\int \d^3x \sqrt{\det g(x)}\;R(x) \longrightarrow 
\sum_l a \delta _l= 2\pi a (N_1-3\k N_3),
\eeq
where we have used that for a closed three-dimensional triangulation
\beq{app4}
\sum_l 1= N_1, ~~~~~\sum_l o(l) = 6 N_3.
\eeq
Taking into account that the 3-volume of an equilateral tetrahedron
is given by $a^{3}/6\sqrt{2}$, we obtain the discretized form of
the Euclidean Einstein action \cite{ajl}
\beq{euclact}
S_{\rm E}=
-\frac{a}{4 G}(N_{1}-3\k N_{3})+
 \frac{a^3 \Lambda}{48 \sqrt{2}\pi G} N_3,
\eeq
where in a slight abuse of language we continue to use $G$ and $\Lambda$
to denote the {\it bare} gravitational and cosmological couplings.
We can substitute $N_1$ by the number $N_0$ of vertices, using the
identity $N_1\equ N_3 +N_0$, which can be derived from the vanishing
of the Euler number for any closed 3d manifold,
\beq{app5}
\chi =N_0-N_1+N_2-N_3 =0,
\eeq
together with the relation $N_2=2 N_3$ 
(any triangle is shared by two tetrahedra and 
any tetrahedron has four triangles). 
Substituting this into \rf{euclact}, we obtain the action used in 
Sec.\ 1,
\beq{app6}
S_E = -k_0N_0+k_3N_3,
\eeq
with the coupling constants given by
\beq{app6a}
k_0 = \frac{a}{4G}, ~~~~
k_3= \frac{a^3\Lambda }{48\sqrt{2}\pi G} +\frac{a}{4G} (3\k-1).
\eeq

\vspace{1cm}
In the numerical investigation of Sec.\ 3 we discussed the dependence of the
action on the total number $N_{22}$ of (2,2)-tetrahedra.
This can be made explicit by rewriting $N_0$ as a function of
$N_3$ and $N_{22}$. For periodic boundary conditions
in the $t$-direction, the total numbers of (3,1)- and (1,3)-tetrahedra
are the same, and we have
\beq{app2}
N_{13}=N_{31}=\frac{1}{2} (N_{3}-N_{22}).
\eeq
Next, we need some identities for the spatial slices at constant
integer $t$. Because the slices are topologically two-spheres, 
the number of vertices in a slice is
\beq{app7}
N_0(t)= \oh N_{31}(t)+2= \oh N_{13}(t-1) + 2.
\eeq
Summing this equation over all $t$ and using (\ref{app2}) yields
\beq{app8}
N_0 =\sum_{t= 1}^T N_0(t) = 2T + \oq (N_3 -N_{22}),
\eeq
and therefore
\beq{app10}
S_E = \frac{k_0}{4} N_{22} + (k_3-\frac{k_0}{4})N_3 -2 k_0 T,
\eeq
which is the form of the action used in Sec.\ 3.

\section*{Appendix 2}

As discussed in Sec.\ 3, one can describe the 3d Lorentzian geometries
in terms of dual graphs, naturally associated with each plane of
constant half-integer $t$. They decompose into two cubic graphs of
different colour (whose trivalent vertices correspond to the (1,3)-
and (3,1)-tetrahedra of the original triangulation, and which may
be thought of as the in- and out-states of the transfer matrix. 
Red and blue lines cross at four-valent vertices, corresponding
to the (2,2)-tetrahedra of the original lattice.

The planarity of this structure (i.e. the fact that the subgraphs
have topology $S^{2}$) is easily implemented in the
program by representing the one-dimensional lines of the red and
blue graphs as double lines with opposite 
orientation, as one can do in the large-$n$ matrix model \rf{matrix}. 
In this way one obtains closed loops of oriented coloured (single) lines 
which are dual to the vertices at times $t$ and $t\pl 
1$ of the original lattice. 
The 2d spherical surface may thus be thought of as being covered by 
(either red or blue) patches enclosed inside the loops.

In the numerical simulations we take care that the triangulations
are 3d simplicial manifolds to start with, and we accept only Monte Carlo
moves which preserve this property. In terms of the original 
triangulation, this means that we only allow the creation of
configurations where 
any two vertices can be shared by
at most one link, any three vertices can be shared by at most one
triangle, and any four vertices by at most one tetrahedron.

In terms of the dual graphs, this implies two types of restrictions,
the first of which have a transparent interpretation in the matrix 
model: they constrain
the individual trivalent graphs to have neither tadpoles nor 
self-energy subdiagrams. This ensures that they are regular 2d
simplicial manifolds with spherical topology. 
The remaining constraints restrict the ways in which the two
coloured graphs are allowed to intersect each other. 
Requiring the absence of double links between pairs of vertices that
are time-like separated on the original lattice implies that 
the intersection of any pair of red and blue domains enclosed by red 
and blue loops cannot be multiply connected. 
Similarly the absence of double triangles from the original
simplicial configuration means that the (one-dimensional) intersection 
of a double line of one colour with a given domain inside a loop 
of the opposite colour must be either empty or simply connected. 

An important consequence of these constraints is 
that the number $N_{22}(t)$ of dual four-valent vertices
is constrained both from below
and above in terms of $N_{13}(t)$ and $N_{31}(t)$. This does not 
happen in the matrix model \rf{matrix}, where these numbers are
completely independent. --
It is possible that some of the regularity conditions discussed here 
can be relaxed without
affecting the universal properties, but for the 3d Lorentzian model
this question has not yet been explored.

\section*{Appendix 3}

The numerical simulations presented in this paper were performed for
system sizes of 4k, 8k, 16k, 32k and 64k tetrahedra, and for total
proper times $T$=16, 32 and 64. As usual, the standard unit was 
taken to be one sweep of the system, interpreted as $N_3$ {\it attempted} 
moves.
Since the acceptance of moves is a function of $k_0$, in order that all 
moves were performed approximately the same number of times, we had to 
tune the number of attempted moves for each of the three types of moves
appropriately.
This technique has been applied successfully before 
in three and four-dimensional simulations
of Euclidean dynamical triangulations.
We considered gravitational couplings in the range between
$k_0=2.0$ and $k_0=7.0$. 
In this range the acceptance of move 1 is between 13.0\% and 75\%, 
the acceptance of the moves 2\&3 between 3\% and 11\% and 
that of the moves 4\&5 between 16\% and 10\%.
For even smaller $k_0$ the acceptance of the moves 2\&3 decreases rapidly 
and it becomes very difficult with the present set-up to change 
the geometry of the spatial intersections. 
A typical run corresponded
to $10^6$ sweeps at a given value of $k_0$. For all measured quantities
we found autocorrelation
times below 100 sweeps, which was also the typical time between successive
measurements.

\end{document}